\documentstyle [12pt,a4,epsf] {article}

\newcommand{\rr   }   { {\bf r} }
\newcommand{\dr  }   { {\rm d}\rr\ }
\newcommand{\half}   {\frac{1}{2}}
\newcommand{\asix}   {\frac{a^2}{6}}
\newcommand{\e   }   { {\rm e} }
\newcommand {\s}{{\rm salt}}
\newcommand{\ie} {{\it i.e.,}}
\newcommand{\eg} {{\it e.g.,}}
\newcommand{\be} {\begin{equation}}
\newcommand{\ee} {\end{equation}}
\newcommand{\bea} {\begin{eqnarray}}
\newcommand{\eea} {\end{eqnarray}}

\begin{document}


\title {Polyelectrolytes in Solution and at
         Surfaces\thanks{to appear in {\it Encyclopedia of Electrochemistry},
          Eds. M. Urbakh and E. Giladi,
         Vol. I, Wiley-VCH, Weinheim, 2002} }
\author {Roland R. Netz \\
         Max-Planck Institute for Colloids and Interfaces\\
         D-14424 Potsdam, Germany \\
\and
David Andelman \\
         School of Physics and Astronomy   \\
         Raymond and Beverly Sackler Faculty of Exact Sciences \\
         Tel Aviv University, Ramat Aviv, Tel Aviv 69978, Israel \\
         \\}
\date{September 2001}
\maketitle

\begin{abstract}
\setlength {\baselineskip} {14pt}

This chapter deals with charged polymers (polyelectrolytes) in
solution and at surfaces. The behavior of polyelectrolytes is
markedly different from that of neutral polymers. In bulk
solutions, i.e. disregarding the surface effect, there are two
unique features to charged polymers: first, due to the presence of
long-ranged electrostatic repulsion between charged monomers, the
polymer conformations are much more extended, giving rise to a
very small overlap concentration and high solution viscosity.
Second, the presence of a large number of counter-ions increases
the osmotic pressure of polyelectrolyte solutions, making such
polymers water soluble as is of great importance to many
applications. At surfaces, the same interplay between
monomer-monomer repulsion and counter-ion degrees of freedom
leads to a number of special properties. In particular, the
adsorption behavior depends on both the concentration of polymers
and added salt in the bulk. We first describe the adsorption
behavior of single polyelectrolyte molecules, and discuss the
necessary conditions to obtain an adsorbed layer and characterize
its width. Depending on the stiffness of the polyelectrolyte, the
layer can be flat and compressed or coiled and extended. We then
proceed and discuss the adsorption of polyelectrolytes from
semi-dilute solutions. Mean-field theory profiles of
polyelectrolyte adsorption are calculated as function of surface
charge density (or surface potential), the amount of salt in the
system and the charge fraction on the chains. The phenomenon of
charge inversion is reviewed and its relevance to the formation
of multilayers is explained. The review ends with a short
overview of the behavior of grafted polyelectrolytes.
\end{abstract}

\pagebreak

\section{Introduction}
\setcounter{equation}{0}

Polyelectrolytes (PE) are charged macromolecules which are
extensively studied not only because of their numerous industrial
applications, but also from a pure scientific
interest~\cite{Oosawa}-\cite{barrat1}.
The most
important property of PE's is their water solubility giving rise
to a wide range of non-toxic, environmentally friendly and cheap
formulations. On the theoretical side, PE's combine the  field of
statistical mechanics of charged systems with the field of polymer
science and offer quite a number of surprises and challenges.

The polymers considered in this review are taken as linear and
long polymer chains, containing a certain fraction of electrically
charged monomers. Chemically, this can be achieved, for example,
by substituting neutral monomers with acidic ones. Upon contact
with water, the acidic groups dissociate into positively charged
protons that bind immediately to water molecules, and negatively
charged monomers. Although this process effectively charges the
polymer molecules, the counter-ions make the PE solution
electro-neutral on macroscopic length scales.

The counter-ions are attracted to the charged polymers via
long-ranged Coulomb interactions, but this physical association
typically only leads to rather loosely bound counter-ion clouds
around the PE chains.
Because PE's are present in a background of a polarizable and
diffusive counter-ion cloud, there is a strong influence of the
counter-ion distribution on the PE structure, as will be discussed
at length in this review. Counter-ions contribute significantly
towards bulk properties, such as the osmotic pressure, and their
translational entropy is responsible for the generally good water
solubility of charged polymers. In addition, the statistics of PE
chain conformation is governed by intra-chain Coulombic repulsion
between charged monomers, resulting in a more extended and
swollen conformation of PE's as compared to neutral polymers.

All these factors combined are of great importance when
considering PE adsorption to charged surfaces. We distinguish
between physical adsorption, where chain monomers are attracted
to surfaces via electrostatic  or non-electrostatic interactions,
and chemical adsorption, where a part of the PE (usually the
chain end) is chemically bound (grafted) to the surface. In all
cases, the long-ranged repulsion of the dense layer of adsorbed
PE's and the entropy associated with the counter-ion distribution
are important factors in the theoretical description.

\section{Neutral Polymers in Solution}
\setcounter{equation}{0}

Before reviewing the behavior of charged polymers, let us
describe some of the important ideas underlying the behavior of
neutral polymer chains in solution.

\subsection{Flexible Chain  Statistics}

The chains considered in this review are either flexible or
semi-flexible. The statistical thermodynamics of flexible chains
is well developed and the theoretical concepts can be applied
with a considerable degree of confidence
\cite{erich}-\cite{degennes}.
Long and flexible chains have a large number of conformations, a
fact that plays a crucial role in determining their behavior in
solution. When flexible chains adsorb on surfaces they form a {\em
diffusive} adsorption layer extending away from the surface into
the solution. This is in contrast to semi-flexible or rigid
chains, which can  form dense and compact adsorption layers.

The main parameters used to describe a flexible polymer chain are
the polymerization index $N$, which counts the number of repeat
units or effective monomers along the chain,
and the Kuhn length $a$, being
the  size of one effective monomer or the distance between two neighboring
effective monomers. The effective monomer size ranges from  a few \AA\ for
synthetic polymers to a few nanometers for biopolymers ~\cite{Flory1}.
The effective monomer size $a$ is not to be confused with the actual
size $b$ of one chemical monomer; in general, $a > b$ due to chain
stiffening effects as will be explained in detail later on.
In contrast to other molecules or particles, a polymer chain
contains not only translational and rotational degrees of
freedom, but also a vast number of conformational degrees of
freedom. For typical polymers, different conformations are
produced by torsional rotations of the polymer backbone bonds.

A simple description of flexible chain conformations is achieved
with the freely-jointed chain model where a polymer consisting of
$N+1$ monomers is represented by $N$ bonds defined by bond
vectors ${\bf r}_j$ with $j = 1, \ldots N$. Each bond vector has
a fixed length $|{\bf r}_j| = a$ corresponding to the Kuhn
length, but otherwise is allowed to rotate freely. For the
freely-jointed chain model, the monomer size $b$ equals the
effective monomer size $a$, $b=a$. Fixing one of the chain ends at
the origin, the position of the $(k+1)$-th monomer is given by the
vectorial sum
 \be
  {\bf R}_k = \sum_{j=1}^k {\bf r}_j .
 \ee
Because  two arbitrary bond vectors are uncorrelated in this
simple model, the thermal average over the scalar product of two
different bond vectors vanishes, $\langle {\bf r}_j\cdot {\bf r}_k
\rangle = 0 $ for $j \neq k$, while the mean squared bond vector
length is simply given by $\langle {\bf r}_j^2 \rangle = a^2 $.
It follows then that the mean squared end-to-end radius is
proportional to the number of monomers
 \be
 \label{fjc} \langle {\bf R}_N^2 \rangle = N a^2.
 \ee
The same result is obtained for the mean quadratic displacement
of a freely diffusing particle  and eludes to the same underlying
physical principle, namely the statistics of Markov processes.

In Figure~1a we show a snapshot of a Monte-Carlo simulation of a
freely-jointed chain consisting of 100 monomers each being
represented by a sphere of diameter $b$ (being equal here to $a$,
the effective monomer size). The bar represents a length of $10
b$, which according to Eq.~(\ref{fjc}) is the average distance
between the chain ends. Indeed, the end-to-end radius gives a
good idea of the typical chain radius.

The freely-jointed chain model describes ideal Gaussian chains
and does not account for interactions between monomers which are
not necessarily close neighbors along the backbone. Including
these interactions will give a different scaling behavior for
long polymer chains. The end-to-end radius, $R=\sqrt{\langle
R^2_N\rangle}$, can be written more generally for $N\gg 1$ as
\be R \sim a N^{\nu}.
\ee
For an ideal polymer chain (no interactions between monomers), the
above result implies $\nu = 1/2$. This result holds only for
polymers where the attraction between monomers (as compared with
the monomer-solvent interaction) cancels the steric repulsion
(which is due to the fact that the monomers cannot penetrate each
other). This situation can be achieved in special solvent
conditions called ``theta" solvents.

More generally, polymers in solution can experience three types of
solvent conditions, with theta solvent condition  being
intermediates between ``good" and ``bad" solvent conditions. The
solvent quality depends mainly on the specific chemistry
determining the interaction between the solvent molecules and
monomers. It also can be changed by varying the temperature.

The solvent is called ``good'' when the monomer-solvent
interaction is more favorable than the monomer-monomer one.
Single polymer chains in good solvents have ``swollen'' spatial
configurations, reflecting the effective repulsion between
monomers. For good solvents, the steric repulsion dominates and
the polymer coil takes a  more swollen structure, characterized by
an exponent $\nu \simeq 3/5$~~\cite{Flory1}. This spatial size of
a polymer coil is much smaller than the extended contour length
$L=aN$ but larger than the size of an ideal chain $aN^{1/2}$. The
reason for this peculiar behavior is entropy combined with the
favorable interaction between monomers and solvent molecules in
good solvents.
As we will see below, it is the conformational freedom of polymer
coils that leads to salient differences between polymer and
simple liquid adsorption.

In the opposite case of ``bad'' (sometimes called ``poor") solvent
conditions, the effective interaction between monomers is
attractive, leading to collapse of the chains and to their
precipitation from the solution (phase separation between the
polymer and the solvent). It is clear that in this case, the
polymer size, like any space filling object embedded in three
dimensional space, scales as $N \sim R^3$, yielding $\nu=1/3$.


\subsection{Semi-Flexible Chain Statistics}

Beside neglecting monomer-monomer interaction, the freely-jointed
chain model does not take into account the chain elasticity,
which plays an important role for some polymers, and leads to
more rigid structures. This stiffness can be conveniently
characterized by the persistence length $\ell_0$, defined as the
length over which the tangent vectors at different locations on
the chain are correlated. In other words, the persistence length
gives an estimate for the typical radius of curvature, while
taking into account thermal fluctuations. For synthetic polymers
with trans-cis conformational freedom of the chain backbone, the
stiffness is due to fixed bond angles and hindered rotations
around individual back-bone bonds. This effect is even more
pronounced for polymers with bulky side chains, such as
poly-DADMAC, because of steric constraints, and the persistence
length is of the order of a few nanometers.

 Biopolymers with a more
complex structure on the molecular level tend to be stiffer than
simple synthetic polymers. Some typical persistence lengths
encountered in these systems are $\ell_0 \approx 5$\,mm for
tubulin,
 $\ell_0 \approx 20$\,$\mu$m for actin, and
$\ell_0 \approx 50$\,nm for double-stranded DNA. Because some of
these biopolymers are charged, we will discuss at length the
dependence of the persistence length on the electrostatic
conditions. In some cases the main contribution to the persistence
length comes from the repulsion between charged monomers.


To describe the bending rigidity of neutral polymers, it is easier
to use a continuum model~\cite{Grosberg}, where one neglects the
discrete nature of monomers. The bending energy (rescaled by the
thermal energy, $k_B T$) of a stiff or semi-flexible polymer of
contour length $L$ is given by
 \be 
 \frac{\ell_0}{2} \int_0^L {\rm d} s \; \left( \frac{ {\rm
d}^2 {\bf r}(s)}{{\rm d} s^2} \right)^2~,
 \ee
where ${\rm d}^2 {\bf r}(s)/{\rm d} s^2$ is the local curvature
of the polymer. We assume here that the polymer segments are
non-extendable, \ie\ the tangent vectors are always normalized,
$|{\rm d} {\bf r}(s)/{\rm d} s|=1$. Clearly, this continuum
description will only be good if the persistence length is larger
than the monomer size. The mean-squared end-to-end radius of a
semi-flexible chain is known and reads~\cite{Grosberg}
\begin{equation} \label{Re2}
R^2
=2  \ell_0 L +2 \ell_0^2\left( {\rm e}^{-L/\ell_0}-1\right)~,
\end{equation}
where the persistence length  is $\ell_0$ and the total contour
length of a chain is $L$. Two limiting behaviors  are obtained
for $R$ from Eq.~(\ref{Re2}): for long chains  $L \gg \ell_0$, the
chains behave as  flexible ones, $R^2 \simeq 2 \ell_0 L $; while
for rather short chains, $ L \ll \ell_0$, the chains behave as
rigid-rods, $R \simeq L$. Comparison with the scaling of the
freely-jointed chain model, Eq.~(\ref{fjc}), shows that a
semi-flexible chain can, for $L \gg \ell_0$, be described by a
freely-jointed chain model with an effective Kuhn length of
 \be
  a= 2 \ell_0
\ee
 and an effective number of segments or monomers
 \be
  N = \frac{L}{2 \ell_0}.
 \ee
In this case the Kuhn length takes into account the chain stiffness
and is  independent from the monomer length. This monomer size is
denoted by $b$ whenever there is need to distinguish between the
monomer size $b$ and  the persistence length $\ell_0$ (or Kuhn
length $a$). In Figure~1 we show snapshots taken from a
Monte-Carlo simulation of a semi-flexible chain consisting of 100
polymer beads of diameter $b$. The persistence length is varied
from $\ell_0=2b$ (Figure~1b), over $\ell_0=10b$ (Figure~1c), to
$\ell_0=100b$ (Figure~1d). Comparison with the freely-jointed
chain model (having no persistence length) is given in Figure~1a
($a=b$, $\ell_0=0$). It is seen that as the persistence length is
increased, the chain structure becomes more expanded. The theoretical
prediction for the average
end-to-end radius $R$, Eq.~(\ref{Re2}), is shown as the black bar
on the figure and gives a  good estimate on typical sizes of
semi-flexible polymers.

\section{Properties of Polyelectrolytes in Solution}
\setcounter{equation}{0}

For polyelectrolytes, electrostatic interactions provide the driving force
for their salient features and have to be included in
any theoretical description. The reduced electrostatic interaction
between two point-like charges
 can be written as $ z_1 z_2 v(r)$ where
\begin{equation} \label{intro1}
v (r) = \frac{e^2}{k_B T\varepsilon  r}
\end{equation}
is the Coulomb interaction between two elementary charges, $z_1$
and $z_2$ are the valences (or the reduced charges in units of
the elementary charge $e$), and $\varepsilon$ is the medium
dielectric constant.
Throughout this review, all energies are given in units of the
thermal energy $k_B T$. The interaction depends only on the
distance $r$ between the charges. The total electrostatic energy
of a given distribution of charges is obtained from adding up all
pairwise interactions between charges according to
Eq.~(\ref{intro1}). In principle, the equilibrium behavior of an
ensemble of charged particles (\eg\  a salt solution) follows from
the partition function, \ie\ the weighted sum over all different
microscopic configurations, which
--- via the Boltzmann factor --- depends on the electrostatic
energy of each configuration. In practice, however, this route is
very complicated for several reasons:

 i) The Coulomb interaction, Eq.~(\ref{intro1}), is long-ranged and
 couples
 many charged particles. Electrostatic problems are
typically {\em many-body problems}, even for low densities.

ii) Charged objects in most cases are dissolved in water. Like any
material, water is polarizable and  reacts to the presence of a
charge with polarization charges. In addition, and this is by far
a more important effect, water molecules carry a permanent dipole
moment that partially orients in the vicinity of charged objects.
Note that for water, $\varepsilon \approx 80$, so that
electrostatic interactions and self energies are much weaker in
water than in air (where $\varepsilon \approx 1$) or some other
low-dielectric solvents. Still, the electrostatic interactions
are especially important in polar solvents because in these 
solvents charges dissociate more easily than in unpolar solvents.

iii) In  biological systems and most industrial applications, the
aqueous solution contains mobile salt ions. Salt ions of opposite
charge are drawn to the charged object and form a loosely bound
counter-ion cloud around it. They effectively reduce or {\em
screen} the charge of the object. The effective (screened)
electrostatic interaction between two charges $z_1 e$ and $z_2 e$
in the presence of salt ions and a polarizable solvent can be
written as $z_1 z_2 v_{\rm DH}(r)$, with the Debye-H\"uckel (DH)
potential $v_{\rm DH}(r)$ given on the linear-response-level by
\begin{equation}
 \label{introDH}
v_{\rm DH} (r) = \frac{\ell_B}{r} {\rm e}^{-\kappa r}.
\end{equation}
The Bjerrum length $\ell_B$ is defined as
\begin{equation}
\ell_B = \frac{e^2}{ \varepsilon k_BT}
\end{equation}
and denotes the distance at which the Coulombic interaction
between two unit charges in a dielectric medium is equal to
thermal energy ($k_BT$). It is a measure of the distance below
which the Coulomb energy is strong enough to compete with the
thermal fluctuations; in water at room temperatures, one finds
$\ell_B \approx 0.7$\,nm. The exponential decay is characterized
by the so-called screening length $\kappa^{-1}$, which is related
to the salt concentration $c_\s$ by
\begin{equation}
\kappa^2 = 8 \pi z^2 \ell_B c_\s
\end{equation}
where $z$ denotes the valency of the screening ions. At
physiological conditions the salt concentration is  $c_\s \approx
0.1$\,M and for monovalent ions ($z=1$) this leads to $\kappa^{-1}
\approx 1$\,nm.

The so-called Debye-H\"uckel interaction, Eq.~(\ref{introDH}),
embodies correlation effects due to the long-ranged Coulomb
interactions in a salt solution using linear-response theory.
\footnote{The number of ions which are correlated in a salt
solution with concentration $c_\s$ is of the order of $n \sim
\kappa^{-3} c_\s$, where one employs the screening length
$\kappa^{-1}$ as the scale over which ions are correlated. Using
the definition $\kappa^2 = 8 \pi z^2 \ell_B c_\s$, one obtains $n
\sim (z^2 \ell_B c_\s^{1/3})^{-3/2}$. The average distance
between ions is roughly $r \sim c_\s^{-1/3}$. The typical
electrostatic interaction between two ions in the solution thus
is $U \sim z^2 \ell_B /r \sim z^2 \ell_B c_\s^{1/3}$ and we
obtain $U \sim n^{-2/3}$. Using these scaling arguments one
obtains that either i) many ions are weakly coupled together (\ie\
$n\gg 1$ and $U \ll 1$), or ii) a few ions interact strongly with
each other ( $n \simeq U \simeq 1$). In the first case, and in
the absence of external fields, the approximations leading to the
Debye-H\"uckel approximation, Eq.~(\ref{introDH}), are valid.}
The DH approximation forms a convenient starting point for treating
screening effects, since (owing to its linear character) 
the superposition principle is valid and the electrostatic free energy
is given by a sum over the two-body potential Eq.(\ref{introDH}).
However, we will at various points in this review also discuss 
how to go beyond the DH approximation, for example in the form 
of the non-linear Poisson-Boltzmann theory (see Section \ref{adssemi})
or a box-model for the counterion distribution (see Section \ref{PEbrush}).

\subsection{Isolated Polyelectrolyte Chains}

  We discuss now the
scaling behavior of a single semi-flexible PE  in the bulk,
including chain stiffness and electrostatic repulsion between
monomers. For charged polymers, the effective persistence length
is increased due to electrostatic repulsion between monomers. This
effect modifies considerably not only the PE behavior in solution
but also their adsorption characteristics.

The scaling analysis is a simple extension of previous
calculations for flexible (Gaussian)
PE's~\cite{Khokhlov}-\cite{Netz2}.
The semi-flexible polymer chain is characterized by a bare
persistence length $\ell_0$ and a linear charge density $\tau$.
Using the monomer length $b$ and the fraction of charged monomers
$f$ as parameters, the linear charge density can be expressed as
$\tau = f/b$. Note that in the limit where the persistence length
is small and comparable to a monomer size, only a single length
scale remains, $\ell_0 \simeq a \simeq b$. Many interesting
effects, however, are obtained in the general case treating the
persistence length $\ell_0$ and the monomer size $b$ as two
independent parameters. In the regime where the electrostatic
energy is weak, and for long enough contour length $L$, $L \gg
\ell_0$,  a polymer coil will be formed with a radius $R$
unperturbed by the electrostatic repulsion between monomers.
According to Eq.~(\ref{Re2}) we get $R^2 \simeq 2 \ell_0 L$. To
estimate when the electrostatic interaction will be sufficiently
strong to swell the polymer coil we recall that the electrostatic
energy (rescaled by the thermal energy $k_BT$) of a homogeneously
charged sphere of total charge $Q$ and radius $R$ is
\be
W_{\rm el} = \frac{ 3 \ell_B Q^2}{5 R}.
\ee
The exact charge distribution inside the sphere only changes the
prefactor of order unity and is not important for the scaling
arguments. For a polymer of length $L$ and line charge density
$\tau$ the total charge is $Q= \tau L$. The electrostatic energy
of a (roughly spherical) polymer coil  is then
\be
W_{\rm el} \simeq \ell_B \tau^2 L^{3/2} \ell_0^{-1/2}.
\ee
The polymer length at which the electrostatic self energy is of
order $k_B T$ \ie\ $W_{\rm el} \simeq 1$, follows as
\begin{equation}
 \label{elblob}
  L_{\rm el} \simeq \ell_0 \left( \ell_B \ell_0
\tau^2 \right)^{-2/3}
\end{equation}
and defines the electrostatic blob size or electrostatic polymer
length. We expect a locally crumpled polymer configuration if
$L_{\rm el} > \ell_0$, \ie\  if
\be
\tau \sqrt{\ell_B \ell_0} < 1 ,
\ee
because the electrostatic repulsion between two segments of
length $\ell_0$ is smaller than the thermal energy and is not
sufficient to align the two segments. This is in accord with more
detailed calculations by  Joanny and Barrat~\cite{barrat2}. A
recent general Gaussian variational calculation confirms this
scaling result and in addition yields logarithmic
corrections~\cite{Netz2}. Conversely, for
\be \label{pers}
\tau \sqrt{\ell_B \ell_0} > 1 ,
\ee
electrostatic chain-chain repulsion is already relevant on length
scales comparable to the persistence length. The chain is
expected to have a  conformation characterized by an effective
persistence length $\ell_{\rm eff}$, larger than the bare
persistence length $\ell_0$, \ie\  one expects $\ell_{\rm eff}
>\ell_0$.

This effect is clearly seen in Figure~2, where we show snapshots
of a Monte-Carlo  simulation of a charged chain of 100 monomers
of size $b$ each and bare persistence length $\ell_0/b = 1$ and
several values of $\kappa^{-1}$ and $\tau$.
The number of monomers in an electrostatic blob can be written
according to Eq.~(\ref{elblob}) as 
$ L_{\rm el}/\ell_0 = (\tau^2\ell_B \ell_0 )^{-2/3}$ 
and yields for Figure 2a)  
$L_{\rm el}/\ell_0 = 0.25$, 
b)  $ L_{\rm el}/\ell_0 = 0.63$, 
c)  $ L_{\rm el}/\ell_0 = 1.6$, and 
d)  $ L_{\rm el}/\ell_0 = 4$. Accordingly, in d) the
electrostatic blobs consist of four monomers, and the weakly
charged chain  crumples at small length scales. A typical
linear charge density reached with synthetic  PE's such as 
Polystyrenesulfonate (PSS) is one charge per
two carbon bonds (or, equivalently, one charge per monomer), and
it corresponds to $\tau \approx 4$\,nm$^{-1}$.
 Since for these highly flexible synthetic PE's the bare persistence length
 is of the order of the monomer size, $\ell_0 \simeq b$, the typical
charge parameter for a fully charged PE therefore is roughly
$\tau^2 \ell_B \ell_0 \approx 3$
 and is intermediate between Figures 2a) and 2b). Smaller
linear charge densities can always be obtained by replacing some
of the charged monomers on the polymer backbone with neutral ones,
in which case the crumpling observed in Figure~2d) becomes relevant.
Larger bare persistence lengths can be achieved with biopolymers
or synthetic PE's with a conjugated carbon backbone.

The question now arises as to what are the typical chain
conformations at much larger length scales. Clearly, they will be
influenced by the repulsions. Indeed, in the {\em persistent
regime}, obtained for $\tau \sqrt{\ell_B \ell_0} > 1 $, the
polymer remains locally stiff even for contour lengths larger than
 the bare persistence length $\ell_0$ and the
effective persistence length is given by
\begin{equation}
 \label{elleff} \ell_{\rm eff} \simeq \ell_0 + \ell_{\rm OSF}.
\end{equation}
The electrostatic persistence length, first derived by Odijk and
independently by Skolnick and Fixman, reads \cite{Odijk0,Skolnick}
\begin{equation}
\label{OSF} \ell_{\rm OSF} = \frac{\ell_B \tau^2}{4 \kappa^2}
\end{equation}
and is calculated from the electrostatic energy of a slightly bent
polymer using the linearized Debye-H\"uckel approximation,
Eq.~(\ref{introDH}). It is valid  only for polymer conformation
which do not deviate too much from the rod-like reference state.
The electrostatic persistence length  gives a sizable
contribution to the effective persistence length only for
$\ell_{\rm OSF} > \ell_0$. This is equivalent to the condition
\be
 \label{Gauss} \tau \sqrt{\ell_B \ell_0} > \ell_0 \kappa.
 \ee
The persistent regime is obtained for parameters  satisfying both
conditions (\ref{pers}) and (\ref{Gauss}). Another regime called
the {\em Gaussian regime} is obtained in the opposite limit of
$\tau \sqrt{\ell_B \ell_0} < \ell_0\kappa$.

The electrostatic persistence length is visualized in Figure~3,
where we present snapshots of a Monte-Carlo simulation of a
charged chain consisting of 100 monomers of size $b$. The bare
persistence length was fixed at $\ell_0 =b$, and the
charge-interaction parameter was chosen to be 
$\tau^2 \ell_B \ell_0 =2$, 
close to the typical charge density in experiments on fully charged
synthetic PE's. 
The snapshots correspond to varying screening length
of a) $\kappa^{-1} = \protect \sqrt{2} b$, leading to an
electrostatic contribution to the persistence length of
$\ell_{\rm OSF} = b$, b) $\kappa^{-1} =\protect \sqrt{18} b$, or
$\ell_{\rm OSF} = 9 b$, and c) $\kappa^{-1} =\protect \sqrt{200}
b$, equivalent to $\ell_{\rm OSF} = 100 b$. According to the
simple scaling principle, Eq.~(\ref{elleff}), the effective
persistence length in the snapshots, Figure~3a-c, should be
similar to the bare persistence length in Figure~1b-d, and
indeed, the chain structures in ~3c) and ~1d) are very similar.
Figure~3a and ~1b) are clearly different, although the effective
persistence length should be quite similar, mostly because of
self-avoidance effects which are present in charged chains and
which will be discussed in detail in Section \ref{selfavoid}.

For the case where the polymer crumples on length scales larger
than the bare bending rigidity, \ie\  for $L_{\rm el} > \ell_0$ or
$\tau \sqrt{\ell_B \ell_0} < 1 $, the electrostatic repulsion
between polymer segments is not strong enough to prevent crumpling
on length scales comparable to $\ell_0$, but can give rise to a
chain stiffening on larger length scales, as explained by
Khokhlov and Khachaturian~\cite{Khokhlov} and confirmed by
Gaussian variational methods~\cite{Netz2}. Figure~4 schematically
shows the PE structure in this regime, where the chain on small
scales consists of Gaussian blobs of chain length $L_{\rm el}$
within which electrostatic interactions are not important. On
larger length scales electrostatic repulsion leads to a chain
stiffening, so that the PE forms a linear array of electrostatic
blobs. To quantify this effect, one defines an effective line
charge density of a linear array of electrostatic blobs with blob
size $R_{\rm el} \simeq \sqrt{\ell_0 L_{\rm el}}$,
\begin{equation}
\label{tautilde}
\tilde{\tau} \simeq \frac{\tau L_{\rm el}}{R_{\rm el}} \simeq
\tau \left( \frac{L_{\rm el}}{\ell_0} \right)^{1/2}
\end{equation}
Combining Eqs.~(\ref{tautilde}) and (\ref{OSF}) gives the
effective electrostatic persistence length for a string of
electrostatic blobs,
\begin{equation}
\label{OSFtilde}
\ell_{\rm KK} \simeq \frac{\ell_B^{1/3} \ell_0^{-2/3}
\tau^{2/3}}{\kappa^2}.
\end{equation}
This electrostatic stiffening is only relevant for the so-called
{\em Gaussian--persistent regime} valid for $\ell_{\rm KK} >
R_{\rm el}$, or equivalently
\be \label{Gausspers} \tau \sqrt{\ell_B \ell_0} >( \ell_0
\kappa)^{3/2}. \ee
When this inequality is inverted the Gaussian persistence regime
crosses over to the Gaussian one.

The crossover boundaries (\ref{pers}), (\ref{Gauss}),
(\ref{Gausspers}) between the various scaling regimes are
summarized in Figure~5. We obtain three distinct regimes. In the
persistent regime, for $\tau \sqrt{\ell_B \ell_0} > \ell_0
\kappa$ and $\tau \sqrt{\ell_B \ell_0} >1 $, the polymer takes on
a rod-like structure with an effective  persistence length larger
than the bare persistence length and given by the OSF expression,
Eq.~(\ref{OSF}). In the Gaussian-persistent regime, for $\tau
\sqrt{\ell_B \ell_0} <1 $ and $\tau \sqrt{\ell_B \ell_0} >
(\ell_0 \kappa)^{3/2}$, the polymer consists of a linear array of
Gaussian electrostatic blobs, as shown in Figure~4, with an
effective persistence length $\ell_{\rm KK}$ larger than the
electrostatic blob size and given by Eq.~(\ref{OSFtilde}).
Finally, in the Gaussian regime, for $\tau \sqrt{\ell_B \ell_0}
<(\ell_0 \kappa)^{3/2} $ and $\tau \sqrt{\ell_B \ell_0} < \ell_0
\kappa$, the electrostatic repulsion does not lead to stiffening
effects at any length scale.

The persistence length $\ell_{\rm KK}$ was also
obtained from Monte-Carlo simulations with parameters shown in
Figure~2d), where chain crumpling at small length scales and
chain stiffening at large length scales occur 
simultaneously\cite{Sim1,Sim2,Sim3,Sim4}.
However, extremely long chains are needed in order to
obtain reliable results for the persistence length, since the stiffening 
occurs only at intermediate length scales and therefore fitting of the 
tangent-tangent correlation function is nontrivial. Nevertheless,
simulations point to a different scaling than in Eq.(\ref{OSFtilde}),
with a dependence on the screening length closer to a linear one,
in qualitative agreement with experimental results\cite{Foerster}.
The situation is complicated by the fact that more recent theories
for the single polyelectrolyte chain give different results,
some confirming the simple scaling results described in Eqs.(\ref{OSF}) and
(\ref{OSFtilde}),\cite{Netz2,Li,Ha2}, some confirming Eq.(\ref{OSF}) while
criticizing Eq.(\ref{OSFtilde})\cite{barrat2,Ha1,Liverpool}.
This issue is not resolved and under intense current investigation.
For multivalent counterions fluctuation effects 
can even give rise to a polyelectrolyte collapse  purely due
to electrostatic interactions\cite{PEcoll1,PEcoll2,PEcoll3}, 
which is accompanied by a negative
contribution to the effective persistence length\cite{PEcoll4}.

\subsubsection{Manning Condensation}

A peculiar phenomenon occurs for highly charged PE's and is known
as the Manning condensation of counter-ions~\cite{Man1,Man2}. For
a rigid PE represented by an infinitely long and straight cylinder
with a linear charge density larger than
\be \label{Manning}
\ell_B \tau z =1
\ee
where $z$ is the counter-ion valency, it was shown that
counter-ions condense on the oppositely charged cylinder even in
the limit of infinite solvent dilution. Real polymers have a
finite length, and are neither completely straight nor in the
infinite dilution limit\cite{Man4,Deserno}. 
Still, Manning condensation has an
experimental significance for polymer solutions because
thermodynamic quantities, such as counter-ion
activities~\cite{Wandrey} and osmotic coefficients~\cite{Blaul},
show a pronounced signature of Manning condensation. Locally,
polymer segments can be considered as straight over length scales
comparable to the persistence length. The Manning condition
Eq.~(\ref{Manning}) usually denotes a region where the binding of
counter-ions to charged chain sections begins to deplete the
solution from free counter-ions. Within the scaling diagram of
Figure~5, the Manning threshold (denoted by a vertical broken
line) is reached typically for charge densities larger than the
one needed to straighten out the chain. This holds for monovalent
ions provided $\ell_0 > \ell_B$, as is almost always the case.
The Manning condensation of counter-ions will not have a profound
influence on the local chain structure since the chain is rather
straight already due to monomer-monomer repulsion. A more
complete description of various scaling regimes related to
Manning condensation, chain collapse and chain swelling has
recently been given\cite{Schiessel}.

\subsubsection{Self-Avoidance and Polyelectrolyte
Chain Conformations } \label{selfavoid}

Let us now consider how the self-avoidance of PE chains comes
into play, concentrating on the persistent regime defined by  $\tau
\sqrt{\ell_B \ell_0} >1 $. The end-to-end radius $R$ of a
strongly charged PE chain shows three distinct scaling ranges.
For a chain length $L$ smaller than the effective persistence
length $\ell_{\rm eff}$, which according to Eq.~(\ref{elleff}) is
the sum of the bare and electrostatic persistence lengths,
$R$ grows linearly
with the length, $R \sim L$. Self-avoidance plays no role in this
case, because the chain is too short to fold back on itself.

For much longer chains, $L \gg \ell_{\rm eff}$, we envision a
single polymer coil as a solution of separate polymer pieces of
length $\ell_{\rm eff}$, and treat their interactions using a
virial expansion.  The second virial coefficient $v_2$ of a rod of
length $\ell_{\rm eff}$ and diameter $d$ scales as $v_2 \sim
\ell_{\rm eff}^2 d$~\cite{Khokhlov}. The chain connectivity is
taken into account by adding the entropic chain elasticity as a
separate term. The standard Flory theory~\cite{Flory1} for a
semi-flexible chain is based on writing the free energy ${\cal F}$ as a
sum of two terms
\be
{\cal F} \simeq  \frac{R^2}{\ell_{\rm eff} L} +
v_2 R^3 \left( \frac{L/\ell_{\rm eff}}{R^3} \right)^2
\ee
where the first term is the entropic elastic energy associated
with swelling a polymer chain to a radius $R$ and the second term
is the second-virial repulsive energy  proportional to the
coefficient $v_2$ and the segment density squared. It is
integrated over the volume $R^3$. The optimal radius $R$ is
calculated by minimizing  this free energy and gives the swollen
radius
\be \label{swollen}
R \sim (v_2 / \ell_{\rm eff})^{1/5} L^\nu
\ee
with $\nu = 3/5$. This swollen radius is only realized above a minimal
chain length $L > L_{\rm sw} \sim \ell_{\rm eff}^7/v_2^2 \sim \ell_{\rm
eff}^3 / d^2$. For elongated segments with $\ell _{\rm eff} \gg
d$, or, equivalently, for a highly charged PE, we obtain an
intermediate range of chain lengths $\ell_{\rm eff} < L < L_{\rm sw}$
for which the chain is predicted to be Gaussian and the chain
radius scales as
\be \label{Grange}
R \sim \ell_{\rm eff}^{1/2} L^{1/2}.
\ee
For charged chains, the effective rod diameter $d$ is given in low
salt concentrations by the screening length, \ie\   $d \sim
\kappa^{-1}$ plus logarithmic corrections. The condition to have a
Gaussian scaling range, Eq.~(\ref{Grange}), thus becomes $\ell
_{\rm eff} \gg \kappa^{-1}$. Within the persistent and the
Gaussian-persistent scaling regimes depicted in Figure~5 the
effective persistence length is dominated by the electrostatic
contribution and given by Eqs.~(\ref{OSF}) and (\ref{OSFtilde}),
respectively, which in turn are always larger than the screening length
$\kappa^{-1}$. It follows that a Gaussian scaling range,
Eq.~(\ref{Grange}), always exists below the asymptotic swollen
scaling range, Eq.~(\ref{swollen}). This situation is depicted in
Figure~4 for the Gaussian-persistent scaling regime, where the
chain shows two distinct Gaussian scaling ranges at the small and
large length scales. This multihierarchical scaling structure is only
one of the many problems one faces when trying to understand 
the behavior of PE chains, be it experimentally, theoretically, or by
simulations.

A different situation occurs when the polymer backbone is under
bad-solvent conditions, in which case an intricate interplay between
electrostatic chain swelling and short-range collapse occurs\cite{Khokhlov2}.
Quite recently, this interplay was theoretically
shown to lead to a Rayleigh instability
in the form of a necklace structure consisting of compact beads connected by
thin strings\cite{Kantor,Dobrynin2,Lyulin,Micka2}.
Small-angle X-ray scattering on solvophobic polyelectrolytes
in a series of polar organic solvents of various solvent quality
could qualitatively confirm these theoretical predictions\cite{Waigh}.

\subsection{Dilute Polyelectrolyte Solutions}

It is natural to generalize the discussion of single chain
behavior to that of many PE chains  in  dilute solvent
concentrations. The dilute regime is defined by $c_m< c_m^{*}$,
where $c_m$ denotes the monomer concentration (per unit volume)
and $c_m^{*}$ is the concentration where individual chains start
to overlap. Clearly, the overlap concentration is reached when
the average bulk monomer concentration exceeds the monomer
concentration inside a polymer coil. To estimate the overlap
concentration $c_m^*$, we simply note that the average monomer
concentration inside a coil with radius $R \simeq b N^\nu$ is
given by
\be c_m^* \simeq \frac{N}{ R^3} \simeq N^{1-3\nu} b^{-3}. \ee
For ideal chains with $\nu=1/2$ the overlap concentration scales
as $c_m^* \sim N^{-1/2}$ and thus decreases slowly as the
polymerization index $N$ increases. For rigid polymers with $\nu
= 1$ the overlap concentration scales as $c_m^* \sim N^{-2}$ and
decreases strongly as $N$ increases. This means that the dilute
regime for semi-flexible PE chains corresponds to extremely low
monomer concentrations. For example taking a monomer size
$b=0.254$\,nm and a polymerization index of $N = 10^4$, the
overlap concentration becomes $c_m^* \approx 6 \times
10^{-7}$\,nm$^{-3} \approx 10^{-3}$\,mM, which is a very small
concentration.

The osmotic pressure in the dilute regime in the limit $c_m
\rightarrow 0$ is given by
\be
\frac{\Pi}{k_BT} = \frac{f c_m}{z} + \frac{c_m}{N}
\ee
and consists of the ideal pressure of non-interacting counter-ions
(first term) and polymer coils (second term). Note that since the
second term scales as $N^{-1}$, it is quite small for large $N$
and can be neglected. Hence, the main contribution to the osmotic
pressure comes from the counter-ion entropy. This entropic term
explains also why charged polymers can be dissolved  in water
even when their backbone is quite hydrophobic. Precipitation of
the PE chains will also mean that the counter-ions are confined
within the precipitate. The entropy loss associated with this
confinement is too large and keeps the polymers dispersed  in
solution. In contrast, for neutral polymers there are no
counter-ions in solution. Only the second term in the osmotic
pressure exists and contributes to the low osmotic pressure of
these polymer solutions. In addition, this can explain the trend
towards precipitation even for very small attractive interactions
between neutral polymers.

\subsection{Semi-Dilute Polyelectrolyte Solution}

In the semi-dilute concentration regime, $c_m > c_m^*$, different
polymer coils are strongly overlapping, but the polymer solution
is still far from being concentrated. This  means that the volume
fraction of the monomers in solution is much smaller than unity,
$b^3 c_m \ll 1$. In this concentration range, the statistics of
counter-ions and polymer fluctuations are intimately connected.
 One example where this
feature is particularly prominent is furnished by neutron and
X-ray scattering from semi-dilute PE
solutions~\cite{Nierlich1}-\cite{Essafi2}.
The structure factor $S(q)$  shows a pronounced peak, which
results from a competition between the connectivity of polymer
chains and the electrostatic repulsion between charged monomers,
as will be discussed below. An important length scale,
schematically indicated in Figure~6, is the mesh-size or
correlation length $\xi$, which measures the length below which
entanglement effects between different chains are unimportant.
The mesh size can be viewed as the polymer (blob) scale below
which single-chain statistics are valid. A semi-dilute solution
can be thought of being composed of roughly a close-packed array
of polymer blobs of size $\xi$.

The starting point for the present discussion is the screened
interaction between two charges immersed in a semi-dilute PE
solution containing, charged polymers, their counter-ions and,
possibly, additional salt ions. Screening in this case is
produced not only by the ions, but also by the charged chain
segments which can be easily polarized and shield any free
charges.

Using the random-phase approximation (RPA), the effective
Debye-H\"uckel (DH) interaction can be written in Fourier space
as~\cite{Borue,Joanny}
\begin{equation} \label{vRPA}
v_{\rm RPA}(q) = \frac{1+ v_2 c_m S_0(q)}
{c_m f^2  S_0(q) +v_{\rm DH}^{-1}(q)+
v_2 c_m v_{\rm DH}^{-1}(q) S_0(q)}
\end{equation}
recalling that $c_m$ is the average density of monomers in
solution and $f$ is the fraction of charged monomers on each of
the PE chains. The second virial coefficient of monomer-monomer
interactions is $v_2$ and the single-chain form factor
(discussed  below) is denoted by $S_0(q)$. In the case where no
chains are present, $c_m = 0$, the RPA expression reduces to
$v_{\rm RPA}(q) =v_{\rm DH}(q)$, the Fourier-transform of the
Debye-H\"uckel potential of Eq.~(\ref{introDH}), given by
\begin{equation} \label{DH}
v_{\rm DH}(q) = \frac{ 4 \pi \ell_B}{q^2 + \kappa^2}.
\end{equation}
As before, $\kappa^{-1}$ is the DH screening length, which is due
to all mobile ions. We can write $ \kappa^2 = \kappa_\s^2 +4 \pi
z \ell_B f c_m $, where $\kappa^2_\s = 8 \pi z^2 \ell_B c_\s$
describes the screening due to added salt of concentration
$c_\s$, and the second term describes the screening due to the
counter-ions of the PE monomers. Within the same RPA
approximation the monomer-monomer structure factor $S(q)$ of a
polymer solution with monomer density $c_m$ is given
by~\cite{Borue,Joanny}
\begin{equation} \label{Sinv}
S^{-1}(q) =
f^2 v_{\rm DH}(q) + S_0^{-1}(q) /c_m + v_2 .
\end{equation}
 The structure factor (or scattering function)
only depends on the form factor of an isolated, non-interacting
polymer chain, $S_0(q)$, the second virial coefficient, $v_2$,
the fraction $f$ of charged monomers, and the interaction between
monomers, which in the present case is taken to be the
Debye-H\"uckel potential $v_{\rm DH}(q)$. The  structure factor
of a non-interacting semi-flexible polymer is characterized, in
addition to the monomer length $b$, by its persistence length
$\ell_{\rm eff}$. In general, this form factor is a complicated
function which cannot be written down in closed
form~\cite{Cloizeaux,Yoshizaki}. However, one can separate between
three different ranges of wavenumbers $q$, and within each range
the form factor shows a rather simple scaling behavior, namely
\begin{equation}
\label{Sasym}
S^{-1}_0(q) \simeq
     \left\{
\begin{array}{llll}
     & N^{-1}
         & {\rm for} &  q^2 < 6/N b \ell _{\rm eff}  \\
     & q^2 b \ell_{\rm eff}/6
         & {\rm for} &  6/N b \ell _{\rm eff}  < q^2 < 36 /
                      \pi^2 \ell_{\rm eff}^2  \\
     & q b / \pi
         & {\rm for }     &  36 / \pi^2 \ell_{\rm eff}^2  < q^2.   \\
\end{array} \right.
\end{equation}
For small wavenumbers the polymer acts like a point scatterer,
while in the intermediate wavenumber regime the polymer behaves
like a flexible, Gaussian polymer, and for the largest
wavenumbers the polymer can be viewed as a stiff rod.

One of the most interesting features of semi-dilute PE solutions
is the fact that the  structure factor $S(q)$  shows a pronounced peak.
 For weakly charged PE's, the
peak position scales as $q \sim c_m^{1/4}$ with the monomer
density~\cite{Moussaid}, in agreement with the above random-phase
approximation (RPA) results for charged
polymers~\cite{Borue,Joanny}. We now discuss the scaling of the
characteristic scattering peak within the present formalism. The
position of the peak follows from the inverse structure factor,
Eq.~(\ref{Sinv}), via $\partial S^{-1}(q) / \partial q =0$, which
leads to the equation
\begin{equation} \label{qmax}
q^2 + \kappa_\s^2 + 4 \pi z \ell_B f c_m  = \left(\frac{8 \pi  q
\ell_B f^2  c_m } { \partial S_0^{-1}(q) /\partial q
}\right)^{1/2}.
\end{equation}
In principle, there are two distinct scaling behaviors possible
for the peak, depending on whether the chain form factor of
Eq.~(\ref{Sasym}) exhibits flexible-like or rigid-like scaling.
Concentrating now on the flexible case, i.e. the intermediate
$q$-range in Eq.~(\ref{Sasym}), the peak of $S(q)$ scales as
\begin{equation} \label{qflex}
q^* \simeq  \left(
 \frac{24 \pi \ell_B f^2 c_m}{b \ell_{\rm eff}} \right)^{1/4},
\end{equation}
in agreement with experimental results. A peak is only obtained
if the left-hand side of Eq.~(\ref{qmax}) is dominated by the
$q$-dependent part, \ie\  if $(q^*)^2 > \kappa_\s^2 + 4 \pi z \ell_B
f c_m$.

%

In Figure~7a we show density-normalized scattering curves for a PE
solution characterized by the persistence length $\ell_{\rm
eff}=1$\,nm (taken to be constant and thus independent of the
monomer concentration), with monomer length $b=0.38$\,nm (as
appropriate for poly-DADMAC) and charge fraction $f=0.5$ and with
no added salt.
As the monomer density decreases (top to bottom in the figure),
the peak moves to smaller wavenumbers and sharpens, in agreement
with previous implementations of the RPA. In Figure~7b we show the
same data in a different representation. Here we clearly
demonstrate that the large-$q$ region already is dominated by the
$1/q$ behavior of the single-chain structure factor, $S_0(q)$.
Since neutron scattering data easily extend to wavenumbers as high
as $q \sim 5$\,nm$^{-1}$, the stiff-rod like behavior in the high
$q$-limit, exhibited on such a plot, will be important in
interpreting and fitting experimental data even at lower
$q$-values.

In a semi-dilute solution there are three different, and in
principle, independent length scales: The mesh size $\xi$, the
screening length $\kappa^{-1}$, and the persistence length
$\ell_{\rm eff}$. In the absence of added salt, the screening
length scales as
\be \kappa^{-1} \sim \left(z  \ell_B f c_m \right)^{-1/2}. \ee
Assuming that the persistence length is larger or of the same
order of magnitude as the mesh size, as is depicted in Figure~6,
the polymer chains can be thought of straight segments between
different crossing links. Denoting the number of monomers inside
a correlation blob as $g$, this means that $\xi \sim b g$. The
average monomer concentration scales as $c_m \sim g / \xi^3$ ,
from which we conclude that
\be \xi \sim \left( b c_m \right)^{-1/2}. \ee
Finally, the persistence length within a semi-dilute PE solution
can be calculated by considering the electrostatic energy cost
for slightly bending a charged rod. In PE solutions, it is
important to include in addition to the screening by salt ions
also the screening due to charged chain segments. This can be
calculated by using the RPA interaction, Eq.~(\ref{vRPA}). Since
the screening due to polymer chains is scale dependent and
increases for large separations, a $q$-dependent instability is
encountered and leads to  a persistence length~\cite{NetzRPA}
\be \ell_{\rm OSF}^{\rm sd} \sim \left( b c_m \right)^{-1/2} \ee
where the `$sd$' superscript stands for `semi-dilute'. This
result is a  generalization of the OSF result for a single chain
and applies to semi-dilute solutions. Comparing the three lengths,
we see that
\be
\xi \sim \ell_{\rm OSF}^{\rm sd} \sim 
\sqrt{\frac{z \ell_B f }{b}} \kappa^{-1}.
\ee
Since the prefactor $ \sqrt{\ell_B f / b}$ for synthetic
fully charged polymers
is  roughly of order unity, one finds that for salt-free
semi-dilute PE solutions, all three length-scales scale in the
same way with $c_m$, namely as $\sim c_m^{-1/2}$, as is known also
from experiments~\cite{Nierlich1,Nierlich2,Spiteri}
and previous theoretical calculations\cite{PEsemi1,PEsemi2}.
In simulations of many polyelectrolyte chains, the reduction of the
chain size due to screening by PE chains was clearly 
seen\cite{Sim6,Sim7,Sim8,Sim9}.

\section{Adsorption of a Single Polyelectrolyte Chain}
\setcounter{equation}{0}

  After reviewing bulk properties of PE solutions we address
the complete adsorption diagram of a single semi-flexible
polyelectrolyte on an oppositely charged substrate. In contrast
to the adsorption of neutral polymers, the resulting phase
diagram shows a large region where the adsorbed polymer is
flattened out on the substrate and creates a dense adsorption
layer.

Experimentally, the adsorption of charged polymers on charged or
neutral substrates has been characterized as a function of the
polymer charge, chemical composition of the substrate, pH and
ionic strength of the solution~\cite{Cohen1,Cohen2},
as well as the substrate charge
density~\cite{Fang}-\cite{Kerstin}. Repeated adsorption of
anionic and cationic PE's can lead to well characterized
multilayers on planar~\cite{hong}-\cite{caruso2}
and spherical substrates~\cite{donath,caruso}. Theoretically, the
adsorption of PE's  on charged surfaces poses a much more
complicated problem than the corresponding adsorption of neutral
polymers. The adsorption process results from a subtle balance
between electrostatic repulsion between charged monomers, leading
to chain stiffening, and electrostatic attraction between the
substrate and the polymer chain. The adsorption problem has been
treated theoretically employing the uniform expansion
method~\cite{Muthukumar} and various continuous mean-field
theories~\cite{Chatellier}-\cite{JFJ99}.
In all these works, the polymer density is taken to be constant in
directions parallel to the surface.

The adsorption of a single semi-flexible and charged chain on an
oppositely charged plane~\cite{Netz4} can be treated as a
generalization of the adsorption of flexible
polymers~\cite{Borisov}. A PE characterized by the
linear charge density  $\tau$, is subject to an electrostatic
potential created by $\sigma$, the homogeneous surface charge
density (per unit area). Because this potential is attractive for
an oppositely charged substrate, we consider it as the driving
force for the adsorption. More complex interactions are
neglected. They are due to the dielectric discontinuity at the
substrate surface and to the impenetrability of the substrate for
salt ions.
\footnote{An ion in solution has a repulsive interaction from the
surface when the solution dielectric constant is higher than that
of the substrate. This  effect can lead to desorption for highly
charged PE chains. On the contrary, when the substrate is a metal
there is a possibility to induce PE adsorption on non-charged
substrates or on substrates bearing charges of the same sign as
the PE. See ref.~\cite{Netz4} for more details.}

Within the linearized DH theory the electrostatic potential of a
homogeneously charged plane is
\be V_{\rm plane} (x) =  4\pi \ell_B \sigma \kappa^{-1} {\rm
e}^{-\kappa x}. \ee
Assuming that the polymer is adsorbed over a layer of width
$\delta$ smaller than the screening length $\kappa^{-1}$, the
electrostatic attraction force per monomer unit length can be
written as
\begin{equation}
\label{fatt} f_{\rm att}  = -4 \pi \ell_B \sigma \tau
\end{equation}
We neglect non-linear effects due to counter-ion condensation.
They are described by the Gouy-Chapman theory for  counter-ion
distribution close to a charged surface. Although these effects
are clearly important, it is difficult to include them
systematically, and we remain at the linearized Debye-H\"uckel
level.

Because of the confinement in the adsorbed layer, the polymer
feels an entropic repulsion. If the layer thickness $\delta$ is
much smaller than the effective persistence length of the polymer,
$\ell_{\rm eff}$, as depicted in Figure 8a, a new length scale,
the so-called deflection length $\lambda$, enters the description
of the polymer statistics. The deflection length $\lambda$
measures the average distance between two contact points of the
polymer chain with the substrate. As shown by Odijk, the
deflection length scales as $\lambda \sim \delta^{2/3} \ell_{\rm
eff}^{1/3}$ and is larger than the layer thickness $\delta $ but
smaller than the persistence length $ \ell_{\rm
eff}$~\cite{Odijk1}. The entropic repulsion follows in a simple
manner from the deflection length by assuming that the polymer
loses roughly a free energy of one $k_B T$ per deflection length.

On the other hand, if $\delta > \ell_{\rm eff}$, as shown in
Figure 8b, the polymer forms a random coil with many loops within
the adsorbed layer. For a contour length smaller than $L \sim
\delta^2/\ell_{\rm eff}$, the polymer obeys Gaussian statistics
and decorrelates into blobs with an entropic cost of one $k_B T $
per blob. The entropic repulsion force per monomer unit length is
thus ~\cite{Odijk1}
\begin{equation}
\label{frep}
    f_{\rm rep}  \sim
     \left\{ \begin{array}{llll}
     & \delta^{-5/3} \ell_{\rm eff}^{-1/3}
         & {\rm for} &  \delta \ll \ell_{\rm eff}  \\
     & \ell_{\rm eff} \delta^{-3}
         & {\rm for }     & \delta \gg \ell_{\rm eff}   \\
                \end{array} \right.
\end{equation}
where we neglected a logarithmic correction factor which is not
important for the scaling arguments. As shown in the preceding
section, the effective persistence length $\ell_{\rm eff}$
depends on the screening length and the line charge density; in
essence, one has to keep in mind that $\ell_{\rm eff}$ is  larger
than $\ell_0$ for a wide range of parameters because of
electrostatic stiffening effects.
\footnote{The situation is complicated by the fact that the
electrostatic contribution to the persistence length is scale
dependent and decreases as the chain is bent at length scales
smaller than the screening length. This leads to modifications of
the entropic confinement force, Eq.~(\ref{frep}), if the
deflection length is smaller than the screening length. As can be
checked explicitly, all results reported here are not changed by
these modifications.}

The equilibrium layer thickness follows from equating the
attractive and repulsive forces, Eqs.~(\ref{fatt}) and
(\ref{frep}). For rather stiff polymers and   small layer
thickness, $\delta < \kappa^{-1} < \ell_{\rm eff}$, we obtain
\begin{equation}
\label{delta}
\delta \sim \left(
\ell_B \sigma \tau \ell_{\rm eff}^{1/3} \right)^{-3/5}
\end{equation}
For a layer thickness corresponding to the screening length,
$\delta \approx \kappa^{-1}$, scaling arguments predict a rather
abrupt desorption transition~\cite{Netz4}. This is in accord with
previous Transfer-Matrix calculations for a semi-flexible  polymer
bound by short-ranged (square-well) potentials
~\cite{Maggs,Gompper,Bundschuh}. Setting $\delta \sim
\kappa^{-1}$ in Eq.~(\ref{delta}), we obtain an expression for the
adsorption threshold (for $\kappa \ell_{\rm eff} > 1$)
\begin{equation}
\label{strong} \sigma^* \sim \frac{\kappa^{5/3} } {\tau \ell_B
\ell_{\rm eff}^{1/3}  }
\end{equation}
For $\sigma   > \sigma^*$ the polymer is adsorbed and localized
over a layer with a width smaller than the screening length (and
with the condition $\ell_{\rm eff}> \kappa^{-1}$ also satisfying
$\delta < \ell_{\rm eff}$). As $\sigma$ is decreased, the polymer
abruptly desorbs at the threshold $\sigma=\sigma^*$ . In the
Gaussian regime, the effective persistence length $\ell_{\rm eff}
$ is given by the bare persistence length $\ell_0$ and the
desorption threshold is obtained by replacing $\ell_{\rm eff} $
by $\ell_0$ in Eq.~(\ref{strong}), \ie\
\begin{equation}
\sigma^* \sim \frac{\kappa^{5/3} } {\tau \ell_B \ell_0^{1/3}  }
\end{equation}
In the persistent regime, we have $\ell_{\rm eff} \sim \ell_{\rm
OSF}$ with $\ell_{\rm OSF}$ given by Eq.~(\ref{OSF}). The
adsorption threshold follows from Eq.~(\ref{strong}) as
\begin{equation}
\sigma^* \sim \frac{\kappa^{7/3} } {\tau^{5/3} \ell_B^{4/3}  }
\end{equation}
Finally, in the Gaussian-persistent regime, we have an effective
line charge density from Eq.~(\ref{tautilde}) and a modified
persistence length, Eq.~(\ref{OSFtilde}). For the adsorption
threshold we obtain from Eq.~(\ref{strong})
\begin{equation}
\sigma^* \sim \frac{\kappa^{7/3} \ell_0^{5/9} } {\tau^{5/9}
\ell_B^{7/9}  }
\end{equation}

Let us now consider the opposite limit, $\ell_{\rm eff} <
\kappa^{-1}$.
\footnote{From Eq.~(\ref{delta}) we see that the layer thickness
$\delta$ is of the same order as $\ell_{\rm eff}$ for $ \ell_B
\sigma \tau \ell_{\rm eff}^2 \sim 1$, at which point the
condition $\delta \ll \ell_{\rm eff} $ used in deriving
Eq.~(\ref{delta}) breaks down. }
If the layer thickness is larger than the persistence length but
smaller than the screening length, $\ell_{\rm eff} < \delta <
\kappa^{-1} $, the prediction for $\delta$ obtained from
balancing Eqs.~(\ref{fatt}) and (\ref{frep}) becomes
\begin{equation}
\label{delta2}
\delta \sim \left( \frac{\ell_{\rm eff}}
{\ell_B \sigma \tau} \right)^{1/3}
\end{equation}
From this expression we see that $\delta$ has the same size as
the screening length $\kappa^{-1}$ for
\begin{equation}
 \label{weak}
\sigma^*  \sim \frac{\ell_{\rm eff} \kappa^3} {\tau \ell_B  }
\end{equation}
This in fact denotes the location of a continuous adsorption
transition at which the layer grows to infinity. The scaling
results for the adsorption behavior of a flexible polymer,
Eqs.~(\ref{delta2})-(\ref{weak}), are in agreement  with previous
results~\cite{Muthukumar}.

In Figure 9 we show the desorption transitions and the line at
which the adsorbed layer crosses over from being flat, $\delta <
\ell_{\rm eff}$, to being crumpled or coiled, $\delta > \ell_{\rm
eff}$. The underlying PE behavior in the bulk, as shown in Figure
5, is denoted by broken lines. We obtain two different phase
diagrams, depending on the value of the parameter
\begin{equation}
\Sigma = \sigma \ell_0^{3/2} \ell_B^{1/2}
\end{equation}
For strongly charged surfaces, $\Sigma > 1$, we obtain the phase
diagram as in Figure 9a, and for weakly charged surfaces, $\Sigma
< 1$, as in Figure 9b. We see that strongly charged PE's, obeying
$\tau \sqrt{\ell_0 \ell_B} > 1$, always adsorb in flat layers.
The scaling of the desorption transitions is in general agreement
with recent computer simulations of charged PE's~\cite{Yamakov}.

\subsection{Adsorption on Curved Substrates}

Adsorption of polyelectrolytes on curved substrates is of
importance because PE's are widely used to stabilize  colloidal
suspensions~\cite{NAP83} and to fabricate hollow polymeric
shells~\cite{donath,caruso}. When the curvature of the small
colloidal particles is large enough, it can lead to a much more
pronounced effect for PE adsorption as compared with
neutral-polymer. This is mainly due to the fact that the
electrostatic energy of the adsorbed PE layer depends sensitively
on curvature~\cite{linse,mateescu,Netz5}. Bending a charged
polymer around a small sphere costs a large amount of
electrostatic energy, which will disfavor adsorption of long,
strongly charged PE at too low salt concentration.

In Figure~10 we show the adsorption phase diagram of a single
stiff PE  of finite length which interacts with an oppositely
charged sphere of charge $Z$ (in units of $e$). The specific
parameters were chosen as appropriate for the complexation of DNA
(a negatively charged, relatively stiff biopolymer) with
positively charged histone proteins, corresponding to a DNA
length of $L = 50$\,nm, a chain persistence length of $\ell_0 =
30$\,nm, and a sphere radius of $R_p=5$\,nm. The phase diagram was
obtained by minimization of the total energy including bending
energy of the DNA, electrostatic attraction between the sphere
and the DNA, and electrostatic repulsion between the DNA
segments with respect to the chain configuration~\cite{Kunze}. 
Configurational fluctuations away from this ground state 
are unimportant for such stiff polymers.

We show in Figure~10 the main transition between an unwrapped
state, at low sphere charge $Z$, and the wrapped state, at large
sphere charge $Z$. It is seen that at values of the sphere charge
between $Z=10$ and $Z=130$ the wrapping only occurs at
intermediate values of the inverse screening length $\kappa\sim
c_\s^{1/2}$. At low salt concentrations, (lower left corner in the phase
diagram), the self-repulsion between DNA segments prevents
wrapping, while at large salt concentrations, (lower right corner in
the diagram), the electrostatic attraction is not strong enough to
overcome the mechanical bending energy of the DNA molecule. These results
are in good agreement with experiments on DNA/histone
complexes~\cite{Yager}. Interestingly, the optimal salt
concentration, where a minimal sphere charge is needed to wrap
the DNA, occurs at physiological salt concentration, for
$\kappa^{-1} \approx 1$\,nm. For colloidal particles of larger
size and for flexible synthetic polymers, configurational fluctuations
become important. They  have been treated using a mean-field
description in terms of the average monomer density profile 
around the sphere~\cite{Goeler,sens}.

\section{Adsorption from Semi-Dilute Solutions} \label{adssemi}
\setcounter{equation}{0}

So far we have been reviewing the behavior of single PE chains
close to a charged wall (or surface). This will be now extended to
include adsorption of PE from bulk (semi-dilute) solutions having
a bulk concentration $c_m^b$. As before the chains are assumed to
have a fraction $f$ of charged monomers, each carrying a charge
$e$, resulting in a linear charge density $\tau=f/b$ on the
chain. The solution can also contain salt (small ions) of
concentration $c_\s$ which is directly related to the
Debye-H\"uckel screening length, $\kappa^{-1}$. For clarity
purposes, the salt is assumed  to be monovalent ($z=1$) throughout
Section 5.

We will consider  adsorption only onto a single flat and charged
surface. Clearly the most important quantity is the profile of
the polymer concentration $c_m(x)$ as function of $x$, the
distance from the wall. Another useful quantity is the polymer
surface excess per unit area,  defined as

\begin{equation}
\Gamma= \int_0^\infty [c_m(x)-c_m^b] dx ~.
\end{equation}
Related to the surface excess $\Gamma$ is the amount of charges
(in units of $e$) carried by the adsorbing PE chains, $f\Gamma$.
In some cases the polymer carries a higher charge (per unit area)
than the charged surface itself, $f\Gamma> \sigma$, and the
surface charge is overcompensated by the PE as we will see later.
This does not violate charge neutrality in the system because of
the presence of the counter-ions in solution.

In many experiments, the total amount of polymer  surface excess
$\Gamma$ is measured as a function of the bulk polymer
concentration, pH and/or ionic strength of the bulk solution
~\cite{Peyser}-\cite{Hoogeveen}.
For reviews see, \eg\
refs.~\cite{Cohen1,Cohen2,FleerBook,Norde}. More recently,
spectroscopy ~\cite{Meadows} and ellipsometry ~\cite{Shubin} have
been used to measure the width of the adsorbed PE layer. Other
techniques  such as neutron scattering can be employed to measure
the entire profile $c_m(x)$ of the adsorbed layer
~\cite{Auvray,Guiselin}.

 In spite of the difficulties
 to treat  theoretically PE's in solution
because of the delicate interplay between the chain connectivity
and the long range nature of electrostatic interactions
~\cite{Oosawa,degennes,Odijk2,Dobrynin}, several simple
approaches treating adsorption exist.
One approach is a discrete {\em multi--Stern layer} model
~\cite{vanderSchee}-\cite{Linse},
 where
the system is placed on a lattice whose sites can be occupied by
a monomer, a solvent molecule or a small ion. The electrostatic
potential is determined self--consistently (mean-field
approximation) together with concentration profiles of the
polymer and small ions.
   In another approach,  the electrostatic potential
and the PE concentration are treated as continuous functions
~\cite{Muthukumar,Varoqui93,epl95,Varoqui91,epr98,mm98,jpc99}.
These quantities are obtained from two coupled differential
equations derived from the total free energy of the system. We
will review the main results of the latter approach, presenting
numerical solutions as well as scaling arguments of the mean field
profiles.

\subsection{Mean-Field Theory and Its Profile Equations}

The charge density on the polymer chains is assumed to be
continuous and uniformly distributed along the chains. Further
treatments of the polymer charge distribution
 (annealed and quenched models)
can be found in refs. ~\cite{epl95,epr98}.
   Within mean--field approximation, the free energy of the
system can be expressed in terms of the local electrostatic potential
 $\psi(\rr)$, the local monomer concentration $c_m(\rr)$
and the local concentration of positive and negative ions
 $c^{\pm}(\rr)$. The mean-field approximation means that the
 influence of the charged surface and the inter-chain interactions
 can be expressed in term of an external potential which will
 determine the local concentration of the monomers, $c_m(\rr)$.
 This external potential depends both on the electrostatic
 potential and on the excluded volume interactions between the
 monomers and the solvent molecules.
The excess free energy with respect to the bulk can then be
calculated using another important approximation, the ground
state dominance. This approximation is used often for neutral
polymers ~\cite{degennes} and is valid for very long polymer
chains, $N\gg 1$. It is then convenient to introduce the polymer
order parameter $\phi(\rr)$ where $c_m(\rr)=|\phi(\rr)|^2$ and to
express the adsorption free energy ${\cal F}$ in terms of $\phi$
and $\psi$ (and in units of $k_BT$)
~\cite{Varoqui93,epl95,Varoqui91,epr98,mm98}
\begin{eqnarray}
   {\cal F} &=& \int \dr \left\{ F_{\rm pol}(\rr) + F_{\rm ions}(\rr)
                        + F_{\rm el}(\rr) \right\}~.
   \label{F}
\end{eqnarray}
  The polymer contribution is
\begin{eqnarray}
   F_{\rm pol}(\rr)  &=&  \asix|\nabla\phi|^2
                 + \half v_2 (\phi^4-\phi_b^4)
                 - \mu_p (\phi^2-\phi_b^2)~,
        \label{fpol}
\end{eqnarray}
where the first term is the polymer elastic energy. Throughout
this section we restrict ourselves to flexible chains and treat
the Kuhn length $a$ and the effective monomer size $b$ as the
same parameter. The second term is the excluded volume
contribution where the second virial coefficient $v_2$ is of order
$a^3$. The last term couples the system to a polymer reservoir
via a chemical potential  $\mu_p$, and $\phi_b=\sqrt{c_m^b}$ is
related to the bulk monomer concentration, $c_m^b$.

   The entropic contribution of the small (monovalent) ions is
\begin{eqnarray}
   F_{\rm ions}(\rr) &=& \sum_{i=\pm}
                   \left[ c^i\ln c^i - c^i
                        - c_\s\ln c_\s + c_\s \right]
                 - \mu^i (c^i-c_\s)~,
        \label{fent}
\end{eqnarray}
   where $c^i(\rr)$ and $\mu^i$ are, respectively,
the local concentration and the chemical potential of the $i=\pm$
ions, while $c_\s$ is the bulk concentration of salt.

Finally, the electrostatic contributions (per $k_BT$) are
\begin{eqnarray}
   F_{\rm el}(\rr)  &=&  \left[ f e\phi^2\psi +e c^+\psi -e c^-\psi
                 - \frac{\varepsilon}{8\pi}|\nabla \psi|^2\right]/k_BT~.
        \label{fel}
\end{eqnarray}
  The first three terms are the
electrostatic energies of the monomers,
the positive ions and the negative ions, respectively,
 $f$ is the fractional charge carried by one monomer.
  The last term is the self energy of the electric field where
 $\varepsilon$ is the dielectric constant of the solution.
Note that the electrostatic contribution, Eq.~(\ref{fel}), is
equivalent to the well known result: $(\varepsilon/8\pi k_BT) \int
\dr |\nabla \psi|^2$ plus surface terms. This can be seen by
substituting  the Poisson--Boltzmann equation (as obtained below)
into Eq.~(\ref{fel}) and then integrating by parts.

Minimization of the free energy, Eqs.~(\ref{F})-(\ref{fel}) with
respect to $c^\pm$, $\phi$ and $\psi$ yields a Boltzmann
distribution for the density of the small ions,
 $c^\pm(\rr)=c_\s \exp(\mp e\psi/k_B T)$, and two
coupled
differential equations for $\phi$ and $\psi$:
\begin{eqnarray}
   \nabla^2\psi(\rr)
   &=& \frac{8\pi e}{\varepsilon}c_\s \sinh(e\psi/k_B T)
    -  \frac{4\pi e}{\varepsilon}
       \left( f\phi^2 - f\phi_b^2 \e^{e\psi/k_B T} \right)~,
   \label{PBs} \\
   \asix \nabla^2\phi(\rr)
   &=& v_2(\phi^3-\phi_b^2\phi) + f\phi e\psi/k_B T~,
   \label{SCFs}
\end{eqnarray}
   Equation~(\ref{PBs}) is a generalized
Poisson--Boltzmann equation including the free ions as well as
the charged polymers. The first term represents the salt
contribution and the second term is due to the charged monomers
and their counter-ions.
 Equation~(\ref{SCFs}) is a generalization of the self--consistent
field equation of neutral polymers ~\cite{degennes}. In the bulk,
the above equations are satisfied by setting $\psi\to 0$ and
$\phi\to \phi_b$.

\subsection{Numerical Profiles: Constant $\psi_s$}

  When the surface is taken as
ideal, {\it \ie}, flat and homogeneous, the physical quantities
depend only on the distance $x$ from the surface. The surface
imposes boundary conditions on the polymer order parameter
$\phi(x)$ and electrostatic potential $\psi(x)$. In thermodynamic
equilibrium all charge carriers in solution should exactly
balance the surface charges (charge neutrality). The
Poisson--Boltzmann equation~(\ref{PBs}), the self--consistent
field equation~(\ref{SCFs}) and the boundary conditions uniquely
determine the polymer concentration profile and the electrostatic
potential. In most cases, these two coupled non--linear equations
can only be solved  numerically.

We present now numerical profiles obtained for surfaces with a
constant potential $\psi_s$:
\begin{eqnarray}
  \left. \psi \right|_{x=0} = \psi_s ~.
  \label{BCpsi0}
\end{eqnarray}
The boundary conditions for $\phi(x)$ depend on the nature of the
short range non-electrostatic interactions of the monomers and the
surface. For simplicity, we  take a non--adsorbing surface and
require that the monomer concentration will  vanish there:
\begin{eqnarray}
  \left. \phi \right|_{x=0} = 0 ~.
  \label{BCphi0}
\end{eqnarray}
We note that the boundary conditions chosen in
Eqs.~(\ref{BCpsi0})-(\ref{BCphi0}) model the particular situation of
electrostatic attraction in competition
with a short range (steric) repulsion of non-electrostatic origin.
  Possible variations of these boundary conditions
include surfaces with a constant surface charge (discussed below)
and surfaces with a non-electrostatic short range attractive (or
repulsive) interaction with  the polymer ~\cite{JFJ99,DAJFJ00}.
 Far from the surface
($x\rightarrow\infty$) both $\psi $ and $\phi $ reach their bulk
values and their derivatives vanish:
 $ \psi'|_{x\rightarrow\infty} = 0$ and
 $\phi'|_{x\rightarrow\infty} = 0$.

The numerical solutions of the mean--field equations~(\ref{PBs}),
(\ref{SCFs}) together with the boundary conditions discussed
above are presented in Figure~11, for several different physical
parameters.
  The polymer is positively charged and is attracted to
the non-adsorbing surface
 held at a constant negative potential.
The aqueous solution contains a small amount of monovalent salt
($c_\s=0.1$\,mM). The reduced concentration profile
$c_m(x)/\phi_b^2$ is plotted as a function of the distance from
the surface $x$.
 Different curves correspond to different values of
the reduced surface potential $y_s\equiv e\psi_s/k_B T$, the
charge fraction $f$  and the effective monomer size $a$.
   Although the spatial variation of the  profiles
differs in detail, they all have a single peak which can be
characterized by its height and width. This observation serves as
a motivation to using scaling arguments.

\subsection{Scaling Results }

The numerical profiles of the previous section indicate that it
may be possible to obtain simple analytical results for the PE
adsorption by assuming that  the adsorption is characterized by
one dominant length scale $D$.  Hence, we write the polymer order
parameter profile in the form
\begin{eqnarray}
   \phi(x)=\sqrt{c_M}h(x/D) ~,
   \label{scaling_prof}
\end{eqnarray}
where $h(x)$ is a dimensionless function normalized to unity at
its maximum and
 $c_M$ sets the scale of polymer adsorption, such that
 $\phi(D)=\sqrt{c_M}$.
The free energy can now be expressed  in terms of $D$ and $c_M$
while the exact form of $h(x)$ affects only the numerical
prefactors.

In principle, the adsorption length $D$ depends also on the ionic
strength through $\kappa^{-1}$. As discussed  below  the scaling
assumption, Eq.~(\ref{scaling_prof}), is only valid as long as
$\kappa^{-1}$ and $D$ are not of the same order of magnitude.
Otherwise, $h$ should be a function of both $\kappa x$ and $x/D$.
We concentrate now on two limiting regimes where
Eq.~(\ref{scaling_prof}) can be justified: (i) the low--salt
regime  $D\ll\kappa^{-1}$ and (ii) the high--salt regime
$D\gg\kappa^{-1}$. We first discuss the case of constant surface
potential which can be directly compared with the numerical
profiles. Then we note the differences with the constant surface
charge boundary condition where the interesting phenomenon of
charge overcompensation is discussed in detail.

\subsubsection{Low--Salt Regime $D\ll\kappa^{-1}$ and $\psi_s=constant$}

In the low--salt regime the effect of the small ions can be
neglected and the free energy (per unit surface area),
Eqs.~(\ref{F})-(\ref{fel}), is approximated by (see also
refs.~~\cite{Varoqui93,mm98})
\begin{eqnarray}
  F \simeq  {a^2\over 6D}c_M -  f|y_s|c_M D
          + 4\pi  l_B f^2 c_M^2 D^3
          + \half  v_2 c_M^2 D ~.
  \label{scalingF}
\end{eqnarray}
In the above equation and in what follows we neglect additional
prefactors of order unity in front of  the various terms which arise from
inserting the scaling profile Eq.(\ref{scaling_prof}) into the 
free energy.
  The first term of Eq.~(\ref{scalingF}) is the elastic energy
characterizing the response of the polymer to concentration
inhomogeneities. The second term accounts for the
electrostatic attraction of the polymers to the charged surface.
 The third term represents the Coulomb repulsion
between adsorbed monomers. The last term represents the excluded
volume repulsion between adsorbed monomers, where we assume that
the monomer concentration near the surface is much larger than
the bulk concentration $c_M\gg\phi_b^2$ .

   In the low--salt regime and for highly charged
PE's the electrostatic interactions are much
stronger than the excluded volume ones.
Neglecting the latter interactions and minimizing the free energy
with respect to $D$ and $c_M$ gives:
\begin{eqnarray}
  D^2 &\simeq&  {a^2 \over f|y_s|}
      \sim {1\over f|\psi_s|}
  \label{scalingD}
\end{eqnarray}
  and
\begin{eqnarray}
  c_M &\simeq&{|y_s|^2 \over 4\pi l_B a^2}
      \sim |\psi_s|^2 ~,
  \label{scalingcm}
\end{eqnarray}
recalling that $y_s=e\psi_s/k_B T$.
As discussed above, these expressions are valid as long as (i)
$D\ll\kappa^{-1}$ and (ii) the excluded volume term in
Eq.~(\ref{scalingF}) is negligible.
 Condition (i) translates into
 $c_\s\ll f |y_s|/(8\pi l_B a^2)$.
For
 $|y_s|\simeq 1$, $a=5$\AA\ and $l_B=7$\AA\
this limits  the salt concentration to
 $c_\s/f \ll $ 0.4 M.
  Condition (ii) on the
magnitude of the excluded volume term
can be shown to be equivalent to
 $f\gg v_2|y_s|/l_B a^2$.
These requirements are consistent with the data presented in
Figure~11.

We recall that the profiles presented in Figure~11 were obtained
from the numerical solution of Eqs.~(\ref{PBs}) and (\ref{SCFs}),
including the effect of small ions and excluded volume.
 The scaling relations
 are verified by plotting in
 Figure~12 the same sets of data as in Figure~11
using rescaled variables as defined in Eqs.~(\ref{scalingD}),
(\ref{scalingcm}). Namely, the rescaled electrostatic potential
 $\psi(x)/\psi_s$ and polymer concentration
 $c_m(x)/c_M\sim c_m(x)a^2/|y_s|^2$ are plotted
as functions of the rescaled distance $x/D\sim x
f^{1/2}|y_s|^{1/2}/a$. The different curves roughly collapse on
the same curve.

  In many experiments the total amount of adsorbed polymer
per unit area $\Gamma$ is measured as function of the physical
characteristics of the system such as the charge fraction $f$,
the pH of the solution or
 the salt concentration $c_\s$
~\cite{Peyser}-\cite{Hoogeveen}.
This quantity can be easily obtained from our scaling expressions
yielding
\begin{eqnarray}
  \Gamma = \int_0^\infty [c_m(x)-\phi_b^2] dx
         \simeq D c_M
         \simeq {|y_s|^{3/2} \over l_B a f^{1/2}}
         \sim {|\psi_s|^{3/2} \over f^{1/2}}~.
 \label{scalingGamma1}
\end{eqnarray}

   The adsorbed amount $\Gamma(f)$ in the low--salt regime
is plotted in the inset of Figure~13a.
 As a consequence of Eq.~(\ref{scalingGamma1}),
 $\Gamma$ decreases with increasing charge fraction $f$.
Similar behavior was also reported in experiments ~\cite{Denoyel}.
This effect is at first glance quite puzzling because as the
polymer charge increases, the chains are subject to a stronger
attraction to the surface. On the other hand, the
monomer--monomer repulsion is stronger and indeed, in this
regime, the  monomer--monomer Coulomb repulsion scales as $(f
c_M)^2$, and dominates over the adsorption energy which scales as
$f c_M$.

\subsubsection{High--Salt Regime $D\gg\kappa^{-1}$ and $\psi_s=constant$}

Let us now consider the opposite case of a high ionic strength
solution. Here $D$ is much larger than $\kappa^{-1}$, and the
electrostatic interactions are short ranged with a cut-off
$\kappa^{-1}$. The free energy of the adsorbing PE layer (per unit
surface area) then reads:
\begin{eqnarray}
  F \simeq {a^2\over 6D}c_M - f|y_s|c_M\kappa^{-1}
          + 4\pi  l_B f^2 \kappa^{-2}c_M^2 D
          + \half  v_2 c_M^2 D~.
  \label{scalingF2}
\end{eqnarray}
The electrostatic cut-off enters in two places. In the second
term only the first layer of width $\kappa^{-1}$ interacts
electrostatically with the surface. In the third term each
charged layer situated at point $x$ interacts only with layers at
$x'$ for which $|x-x'|<\kappa^{-1}$. This term can be also viewed
as an additional electrostatic excluded volume
 with $v_{\rm el} \sim l_B (f/ \kappa)^2$.

   Minimization of the free energy gives
\begin{eqnarray}
  D &\simeq&  {\kappa a^2\over f|y_s|}
     \sim {c_\s^{1/2} \over f|\psi_s|}
  \label{scalingD2}
\end{eqnarray}
  and
\begin{eqnarray}
  c_M &\sim& { f^2 |y_s|^2/(\kappa a)^2 \over
                     f^2/c_\s + \alpha v_2}
  \label{scalingcm2}
\end{eqnarray}
  yielding
\begin{eqnarray}
  \Gamma &\sim& {f |y_s|c_\s^{-1/2} \over
                     f^2/c_\s + \alpha v_2} \sim
{f|\psi_s|\over {v_{\rm el} + \alpha v_2}}c_\s^{-1/2} ~,
  \label{scalingGamma2}
\end{eqnarray}
where $\alpha$ is a numerical constant of order unity which
depends on the profile details.

 The adsorption behavior is depicted in Figures~13 and 14.
Our results are in agreement with numerical solutions of discrete
lattice models  (the multi--Stern layer theory)
~\cite{Cohen1,Cohen2,FleerBook,vanderSchee,Papenhuijzen,Evers,vandeSteegSim,Linse}.
In Figure~13,  $\Gamma$ is plotted as function of $f$ (Figure~13a)
and the pH (Figure~13b) for different salt concentrations.
  The behavior as seen on Figure~13b represents annealed
 PE's
where the nominal charge fraction is given 
by the pH of the solution through the expression
\begin{eqnarray}
  f = {10^{{\rm pH-pK}_0} \over 1 + 10^{{\rm pH-pK}_0} } ~,
  \label{pHpK}
\end{eqnarray}
where ${\rm pK}_0 = -\log_{10}{\rm K}_0$ and ${\rm K}_0$ is the
apparent dissociation constant.
We note that this relation is only strictly valid for 
infinitely dilute monomers and
that distinct deviations from it are observed for polyelectrolytes
because of the electrostatic repulsion between neighboring dissociating
sites\cite{Tanford}. Still, the results in Figure~13b capture the main
qualitative trends of pH-dependent PE adsorption.
Another interesting observation which can be deduced from
Eq.~(\ref{scalingGamma2}) is that $\Gamma$ is only a function  of
 $fc_\s^{-1/2}$. Indeed, as can be seen in Figure~13,
 $c_\s$  only affects the position
of the peak and not its height.

The effect of salt concentration is shown in Figure~14, where
  $\Gamma$ is plotted as function of
the salt concentration $c_\s$ for two charge fractions
 $f=0.01$ and $0.25$.
The curves on the right hand side of the graph are calculated
from the high--salt expression for $\Gamma$,
Eq.~(\ref{scalingGamma2}). The horizontal lines on the left hand
side of the graph indicate the low--salt values of $\Gamma$,
Eq.~(\ref{scalingGamma1}). The dashed lines in the intermediate
salt regime  serve only as guides to the eye since our scaling approach
is not valid when $D$ and $\kappa^{-1}$ are of the same order.

Emphasis should be drawn to the  distinction between weakly and
strongly charged  PE's. For weak PE's, the adsorbed amount $\Gamma$ is a
monotonously decreasing function of the salt concentration $c_\s$
in the whole range of salt concentrations. The reason being that
the monomer--monomer Coulomb repulsion, proportional to $f^2$, is
weaker than the monomer--surface interaction, which is linear in
$f$.

For strongly charged  PE's, on the other hand, the balance between these two
electrostatic terms depends on the amount of salt. At low salt
concentrations, the dominant interaction is the monomer--surface
Coulomb repulsion. Consequently, addition of salt screens this
interaction and increases the adsorbed amount. When the salt
concentration is high enough, this Coulomb repulsion is screened
out and the effect of salt is to weaken the surface attraction.
At this point the adsorbed amount starts to decrease. As a
result, the behavior over the whole concentration range is
non-monotonic with a maximum at some optimal value $c_\s^*$ as
seen in Figure~14.

 From this analysis and from Figures~13,14 and Eq.~(\ref{scalingGamma2}),
it is now natural to divide the high--salt regime into two sub
regimes according to the PE charge. At low charge fractions
(sub-regime HS I),
 $f\ll f^*=(c_\s v_2)^{1/2}$,
the  excluded volume term dominates the denominator of
Eq.~(\ref{scalingGamma2}) and
\begin{eqnarray}
\Gamma &\sim& {f|\psi_s| c_\s^{-1/2}} ~,
  \label{scalingHSI}
\end{eqnarray}
whereas at high $f$ (sub-regime HS II),
 $f\gg f^*$,
the monomer--monomer electrostatic repulsion dominates and
$\Gamma$ decreases with $f$ and increases with $c_\s$:
\begin{eqnarray}
 \Gamma &\sim& {c_\s^{1/2} |\psi_s| f^{-1}} ~.
  \label{scalingHSII}
\end{eqnarray}
The various regimes with their crossover lines are shown
schematically in Figure~15. Keeping the charge fraction $f$
constant while changing the amount of salt corresponds to a
vertical scan through the diagram. For weak PE's this cut goes
through the left hand side of the diagram starting from the
low--salt regime and, upon addition of salt, into the HS I
regime. Such a path describes the monotonous behavior inferred
from Figure~14 for the weak PE ($f=0.01$). For strong PE the cut
goes through the right hand side of the diagram, starting from
the low--salt regime, passing through the HS II regime and ending
in the HS I.
 The passage through the HS II regime is responsible to the
non-monotonous behavior inferred from Figure~14 for the strong PE
($f=0.25$).

Similarly, Figures~13a,13b correspond to horizontal scans through
the top half of the diagram. As long as the system is in the HS I
regime, the adsorbed amount increases when the polymer charge
fraction increase. As the polymer charge further increases, the
system enters the HS II regime and the adsorbed amount decreases.
Thus, the non-monotonous behavior of Figure~13.
We finally note that the single-chain desorption transition for
a flexible chain which was discussed in Section 4, is also 
valid for adsorption from solutions and will lead to a desorption 
transition at very high salt concentrations in Fig.~15.

\subsection{Overcompensation of Surface Charges: Constant $\sigma$}

We turn now  to a different electrostatic boundary condition of
constant surface charge density and look at the interesting
phenomenon of charge compensation by the PE chains in relation to
experiments for PE adsorption on flat surfaces, as well as on
charged colloidal particles~\cite{Decher,donath,caruso}. What was
observed in experiments is that PE's  adsorbing on an oppositely
charged surface can overcompensate the original surface charge.
Because the PE's create a thin layer close to the surface, they can
act as an effective absorbing surface to a second layer of PE's having an
opposite charge compared to the first layer. Repeating the
adsorption of alternating positively and negatively charged PE's,
it is possible to create a multilayer structure of PE's at the
surface. Although many experiments and potential applications for
PE multilayers exist, the  theory of PE overcompensation is only
starting to be
developed~\cite{JFJ99,Netz4,epr98,mm98,jpc99,DAJFJ00,Nguyen}.

The scaling laws presented for constant $\psi_s$ can be used also
for the case of constant surface charge. A surface held at a
constant potential $\psi_s$ will  induce a surface charge density
$\sigma$. The two quantities are related by: $d\psi/dx=4\pi
\sigma e/\varepsilon$ at $x=0$. We will now consider separately the
two limits: low salt $D \ll \kappa^{-1}$, and high salt  $D \gg
\kappa^{-1}$.

\subsubsection{Low Salt Limit: $D\ll \kappa^{-1}$}

Assuming that there is only one length scale characterizing the
potential behavior in the vicinity of the surface, as demonstrated
in Figure~12a, we find that  the surface potential $\psi_s$ and the 
surface charge $\sigma$ are related by $\psi_s \sim \sigma e D$. 
In the low salt limit we find from  Eq.(\ref{scalingD})

\begin{equation}
D \sim (f\sigma l_B)^{-1/3} ~.
\label{Ds}
\end{equation}
Let us define two related concepts via the effective surface
charge density  defined as $\Delta \sigma = f\Gamma - \sigma$,
which is sum of the adsorbed polymer charge density
and the charge density of the bare substrate. 
For $\Delta \sigma =0$ the adsorbed polymer charge exactly
{\it compensates} the substrate charge.
If $\Delta \sigma$ is positive the PE {\it overcompensates} 
the substrate charge, more polymer adsorbs than is needed 
to exactly cancel the substrate charge.
If $\Delta \sigma$
is positive and reaches the value $\Delta
\sigma=\sigma$ it means that the PE charge is $f\Gamma=2\sigma$
and leads to a {\it charge inversion} of the substrate charge. 
The effective surface charge consisting of the substrate charge 
plus the PE layer has a charge density which is exactly opposite to
the original substrate charge density $\sigma$.

Do we obtain overcompensation or
charge inversion in the low salt limit within
mean-field theory? Using scaling arguments this is not clear since
we find that $\Delta \sigma\sim f\Gamma \sim \sigma$. Namely each
of the two terms in $\Delta \sigma$ scales linearly with $\sigma$,
and the occurrence of overcompensation or
charge inversion will depend on numerical prefactors
determining the relative sign of the two opposing terms. However,
if we look on the numerical solution for the mean-field
electrostatic potential, Figure~12, we see indeed that all
plotted profiles have a maximum of $\psi(x)$ as function of $x$.
An extremum in $\psi$ means a zero local electric field. Or
equivalently, using Gauss law, this means that the integrated
charge density from the wall to this special extremum point
(including surface charges) is exactly zero. At this point the
charges in solution exactly compensate the surface charges.

\subsubsection{High Salt Limit: $D\gg \kappa^{-1}$}

 When we include salt in
the solution and look at the high salt limit the situation is
more complex. The only length characterizing the exponential 
decay of $\psi$
close to the surface is the Debye-H\"uckel screening length.
Hence, using $d\psi/d x|_s\sim \sigma e \kappa^{-1}$
yields $\psi_s \sim \sigma e \kappa^{-1}$ and
therefore from Eq.(\ref{scalingD2})

\begin{equation}
D\sim {\kappa^2 a^2 \over f\sigma l_B} \sim \kappa^2 f^{-1}
\sigma ^{-1} \label{Dp}
\end{equation}
The estimation of the PE layer charge can be obtained by using the expression for $D$ and $c_M$
in this high salt limit, Eqs. (\ref{scalingD2})-(\ref{scalingGamma2}),  yielding

\begin{equation}
f\Gamma \simeq \frac{\beta \sigma (8\pi l_B c_\s\kappa^{-2})}{1+\alpha v_2 c_\s f^{-2}}
=\frac{\beta \sigma}{1+v_2/v_{\rm el}}
\label{Dg}
\end{equation}
%
%
where $v_{\rm el}= f^2/\alpha c_{\s}$ is the
electrostatic contribution to the second virial coefficient
$v_2$, and $\alpha>0$ and $\beta>1$ are positive numerical factors.

 We see that $\Delta\sigma=f\Gamma-\sigma$ is a decreasing function of
$v_2$. Charge overcompensation can occur when $v_2$ is smaller than
$v_{\rm el}$ (up to a prefactor of order unity). When $v_2$ can
be neglected in the vicinity of the surface, or when $v_2=0$
(theta solvents), there is always charge overcompensation, $\Delta \sigma=
(\beta-1)\sigma>0$. This is the
case of strongly charged PE. Similar conclusions have been
mentioned in refs.~\cite{JFJ99,DAJFJ00} where $v_2$ was taken as
zero but the surface has a non-electrostatic short range
interaction with the PE. By tuning the relative strength of the
surface charge density $\sigma$ and the non-electrostatic
interaction it is also possible to cause a charge overcompensation and
even an exact charge inversion in a special case.

Finally, we note that the dependence of the charge parameter
$\Delta \sigma$ on the amount of salt, $c_{\s}$, is different for
constant surface charge and constant surface potential cases.
While for the former $\Delta \sigma$ is non-monotonous and has a
maximum (as mentioned above) as function of the salt
concentration, in the latter case $\Delta \sigma/\sigma$ is a
monotonic decreasing function of $c_\s$, Eq.~(\ref{Dg}). This
can be explained by the extra powers of $c_\s$ in the latter case
coming from the relation $\psi_s\sim \sigma c_\s^{-1/2}\sim \sigma\kappa^{-1}$.

Let us remark that in other theories the overcharging is due to
lateral correlations between adsorbed PE's, which in conjunction
with screening by salt ions leads to strongly overcharged
surfaces ~\cite{Netz4,Nguyen}.

\subsection{Final Remarks on Adsorption from Semi-Dilute Solutions}

The results  presented above (Section 5) for adsorption from
solutions have been derived using mean-field theory. Hence,
lateral fluctuations in the polymer and ionic concentrations are
neglected. In addition, we neglect the delicate influence of the
charges on the PE persistence length and any deviations from
ground state dominance ~\cite{degennes}. The region of validity
of the theory is for long and weakly charged polymer chains in
contact with a moderately charged surface. The PE solution is
placed in contact with a single and ideal surface (infinite, flat
and homogeneous). The problem reduces then to an effective one
dimensional problem depending only on the distance from the
charged surface. We take very simple boundary conditions for the
surface assuming that the polymer concentration is zero on the
surface and keeping the surface in constant potential or constant
surface charge conditions.

 We find numerical solutions for the PE profile
 equations in various cases. These numerical solutions agree
well with
 simple  scaling assumptions
describing the adsorption of PE's.  Scaling expressions for the
amount of adsorbed polymer $\Gamma$ and the width $D$ of the
adsorbed layer, as a function of the fractional charge $f$ and
the salt concentration $c_\s$, are obtained for two cases:
constant $\psi_s$ and constant $\sigma$.

For constant $\psi_s$ and   in the low--salt regime a $f^{-1/2}$
dependence of $\Gamma$ is found. It is supported by our numerical
solutions of the profile equations (\ref{PBs}), (\ref{SCFs}) and
is in agreement with experiment ~\cite{Denoyel}.
   This  behavior is due to strong Coulomb repulsion between
adsorbed monomers in the absence of salt. As $f$ decreases, the
adsorbed amount increases until the electrostatic attraction
becomes  weaker than the excluded volume repulsion, at which
point $\Gamma$ starts to decrease rapidly.
 At high salt
concentrations it is not possible to neglect the excluded volume
interaction of the monomers since the electrostatic interactions
are screened by the salt. We obtain two limiting behaviors: (i)
for weakly charged PE's, $f\ll f^*= ({c_\s v_2})^{1/2}$, the
adsorbed amount increases with the fractional charge and
decreases with the salt concentration, $\Gamma\sim
f/\sqrt{c_\s}$, due to the monomer--surface electrostatic
attraction. (ii) For strong PE's, $f\gg f^*$, the adsorbed amount
decreases with the fractional charge and increases with the salt
concentration, $\Gamma\sim \sqrt{c_\s}/f$, due to the dominance
of the monomer--monomer electrostatic repulsion. Between these
two regimes  we find that the adsorbed amount reaches a maximum
in agreement with experiments ~\cite{Blaakmeer,Hoogeveen}.

The scaling arguments are then repeated for constant $\sigma$
boundary conditions. It is found that the PE can possibly  cause
charge overcompensation and even
inversion of the nominal substrate charge, leading the way
to multilayer formation of positively and negatively charged PE.
   The scaling approach
can serve as a starting point for further investigations. The
analytical and approximated expressions are valid only in
specific limits. Special attention should be directed to the
crossover regime where $D$ and $\kappa^{-1}$ are of comparable
size.

The problem of charge inversion is not well understood at
present. Alternative approaches rely on lateral correlations
between semi-flexible adsorbed PE chains which also can lead to
strong overcompensation of surface charges ~\cite{Netz4,Nguyen}.


\section{Polyelectrolyte Brushes} \label{PEbrush}
\setcounter{equation}{0}

Charged polymers which are densely end-grafted to a surface are
called polyelectrolyte  or charged brushes. They have
been the focus of numerous theoretical
~\cite{MIK88}-\cite{ZHU97} and experimental
~\cite{MIR95,GUE95,AHR97} studies. In addition to the basic
interest, charged brushes are considered for their applications
as efficient means for preventing colloids
in polar media (such as aqueous solutions) from flocculating and
precipitating out of solution~\cite{NAP83}. This stabilization
arises from steric (entropic)
as well as electrostatic (energetic) repulsion. A strongly charged brush
is able to trap its own counter-ions and generates a layer of
locally enhanced salt concentration~\cite{PIN91}. It is thus less
sensitive to the salinity of the surrounding aqueous medium than
a stabilization mechanism based on pure electrostatics (\ie\
without polymers).

Neutral brushes have been extensively studied theoretically in
the past using scaling theories~\cite{alex,gennes},
strong-stretching theories~\cite{sem,mil,skvor}, self-consistent
field theories~\cite{netzbrush}, and computer
simulations~\cite{murat,Seidel}. Little is known from experiments
on the scaling behavior of PE brushes as compared to uncharged
polymer brushes. The thickness of the brush layer has been
calculated from neutron-scattering experiments on end-grafted
polymers ~\cite{MIR95} and charged diblock-copolymers at the
air-water interface ~\cite{AHR97}.

Theoretical work on PE brushes was initiated by the works of
Miklavic and Marcelja ~\cite{MIK88} and Misra et al.
~\cite{MIS89}. In 1991, Pincus ~\cite{PIN91} and Borisov,
Birshtein and Zhulina ~\cite{BOR91} presented scaling theories for
charged brushes in the so-called osmotic regime, where the brush
height results from the balance between the chain elasticity
(which tends to decrease the brush height) and the repulsive
osmotic counter-ion pressure (which tends to increase the brush
height). In later studies, these works have been generalized to
poor solvents ~\cite{ROS92} and to the regime where excluded
volume effects become important, the so-called quasi-neutral or
Alexander regime~\cite{BOR94}.

In what follows we assume that the charged brush is characterized
by two length scales: the average vertical extension of polymer
chains from the wall $L$, and the typical extent of the
counter-ion cloud, denoted by $H$. We neglect the presence of
additional salt, which has been discussed extensively in the
original literature, and only consider screening effects due to
the counter-ions of the charged brush. Two different scenarios
emerge, as is schematically presented in Figure~16. The
counter-ions can either extend outside the brush, $H\gg L$, as
shown in 16a), or be confined inside the brush, $H\approx L$, as
shown in 16b). As we show now, case b) is indicative of strongly
charged brushes, while case a) is typical for weakly charged
brushes.

The free energy per unit area (and in units of $k_B T$) contains
several contributions. We denote the grafting density of PE's by
$\rho$, the counter-ion valency by $z$, recalling that $N$ is the
polymerization index of grafted chains and $f$ their charge
fraction.
 The osmotic free energy, $F_{\rm os}$,
associated with the ideal entropy cost of confining the
counter-ions to a layer of thickness $H$ is given by

\begin{equation}
    F_{\rm os}  \simeq
\frac{N f \rho}{z}
              \ln \left( \frac { N f \rho}{z H}\right).
\end{equation}

$F_{\rm v_2}$ is the second virial contribution to the free
energy, arising from steric repulsion between the monomers
(contributions due to counter ions are neglected). Throughout
this review, the polymers are assumed to be in a good solvent
(positive second virial coefficient $v_2 > 0$). The contribution
thus reads
\begin{equation}
  F_{\rm v_2}   \simeq  \frac{1}{2} \ L v_2
\left(
      \frac{N \rho}{L}\right)^2.
\end{equation}
Finally, a direct electrostatic contribution $F_{\rm el}$ occurs
if the PE brush is not locally electro-neutral  throughout the
system, as for example is depicted in Figure~16a. This energy is
given by
\begin{equation}
F_{\rm el} = \frac{2 \pi \ell_{\rm B} (N f \rho)^2}{3}
                \frac{(L-H)^2}{H}.
\end{equation}
This situation arises in the limit of low charge, when the
counter-ion density profile extends beyond the brush layer, \ie\
$H> L$.

The last contribution is the stretching energy of
the chains which is
\begin{equation}
F_{\rm st}  = \frac{3 L^2}{2 N a^2}\, \rho.
\end{equation}
Here, $a$ is the monomer size or Kuhn length of the polymer,
implying that we neglect any chain stiffness for the brush
problem. The different free energy contributions lead, upon
minimization with respect to the two length scales $H$ and $L$,
to different behaviors. Let us first consider the weak charging
limit, \ie\  the situation where the counter-ions leave the brush,
$H > L$. In this case, minimization of $F_{\rm os} + F_{\rm el}$
with respect to the counter-ion height $H$ leads to
 \be
  H \sim \frac{1}{z \ell_B N f \rho}
 \ee
which is the Gouy-Chapman length for $z$-valent counter-ions at a
surface of surface charge density $\sigma = N f \rho$. Balancing
now the polymer stretching energy $F_{\rm st}$ and the
electrostatic energy $F_{\rm el}$ one obtains the so-called
Pincus brush
\begin{equation}
         L \simeq N^3 \rho\, a^2\ell_B f^{2} ~.
         \label{Pinc}
\end{equation}
In the limit of $H\approx L$ 
the PE brush can be considered as neutral and the electrostatic
energy vanishes. There are two ways of balancing the remaining
free energy contributions. The first is obtained by comparing the
osmotic energy of counter-ion confinement, $F_{\rm os}$, with the
polymer stretching term, $F_{\rm st}$, leading to the height
 \be
L \sim \frac{N a f^{1/2}}{z^{1/2}}~,
 \ee
constituting the so-called osmotic brush regime. Finally
comparing the second-virial free energy, $F_{\rm v_2}$, with the
polymer stretching energy, $F_{\rm st}$, one obtains
 \be
  L \sim N a \left( v_2 \rho /a \right)^{1/3}~,
 \ee
and the PE brush is found to have the same scaling behavior as
the neutral brush~\cite{alex,gennes}. Comparing the heights of
all three regimes we arrive at the phase diagram shown in
Figure~17. The three scaling regimes meet at the characteristic
charge fraction
\be
f^* \sim \left( \frac{z v_2}{N^2 a^2 \ell_B} \right)^{1/3}
\ee
and the characteristic grafting density
\be
\rho^* \sim \frac{1}{N \ell_B^{1/2} v_2^{1/2}}.
\ee
For large values of the charge fraction $f$ and the grafting
density $\rho$ it has been found numerically that the brush height
does not follow any of the scaling laws discussed
here~\cite{Csajka0}. This has been recently rationalized in terms
of another scaling regime, the collapsed regime. In this regime
one finds that correlation and fluctuation effects, which are
neglected in the discussion in this section, lead to a net
attraction between charged monomers and
counter-ions~\cite{Csajka}.

\section{Conclusions}
\setcounter{equation}{0}

In this chapter we have reviewed the behavior of charged polymers
(polyelectrolytes) in solution and at interfaces, concentrating on
aspects that are different from the corresponding behavior of
neutral polymers.

Because polyelectrolytes (PE) tend to be quite stiff due to
electrostatic monomer-monomer repulsions, their chain statistics
is related to that of semi-flexible polymers. Neutral and charged
semi-flexible polymers are controlled by their bending rigidity,
which is usually expressed in terms of a persistence length (see
Section 2.2). For PE's, the electrostatic interaction influences
considerably this persistence length.

In solution, we have considered the scaling behavior of a single
PE (Section 3.1). The importance of the electrostatic persistence
length was stressed. The Manning condensation of counter-ions
leads to a reduction of the effective linear charge density
(Section 3.1.1). Excluded volume effects are typically less
important than for neutral polymers (Section 3.1.2). Dilute PE
solutions are typically dominated by the behavior of the
counter-ions. So is the large osmotic pressure of dilute PE
solutions due to the entropic contribution of the counter-ions
(Section 3.2). Semi-dilute PE solutions can be described by the
random-phase-approximation (RPA), which in particular yields the
characteristic peak of the structure factor.

At surfaces, we  discussed in detail the adsorption of single PE
(Section 4), the adsorption from semi-dilute solutions (Section
5), and the behavior of end-grafted PE chains (Section 6). We
tried to express the PE behavior in terms of a few physical
parameters such as the chain characteristics (persistence length),
ionic strength of the solutions and surface characteristics. The
shape and size of the adsorbing layer is, in many instances,
governed by a delicate balance of competing mechanisms of
electrostatic and non-electrostatic origin. In some cases it is
found that the adsorbing PE layer is flat and compressed, while
in other cases it is coiled and extended. Yet, in other
situations, the PE will not adsorb at all and will be depleted
from the surface.  We also briefly review the phenomenon of
charge overcompensation and inversion, 
when the adsorbed PE layer effectively
inverses the sign of the surface charge leading the way to
formation of PE multilayers.

Important topics that we have left out are the dynamics of PE
solutions, which is reviewed in ref.~\cite{barrat1}, and the
behavior of PE under bad-solvent 
conditions~\cite{Dobrynin2,Lyulin,Micka2}. In
the future we expect that studies of PE in solutions and at
surfaces will be directed more towards biological systems. We
mentioned in this review the complexation of DNA and histones
(Sec. 4.1). This is only one of many examples of interest where
charged biopolymers, receptors, proteins and DNA molecules
interact with each other or with other cellular components. The
challenge for future fundamental research will be to try to
understand the role of electrostatic interactions combined with
specific biological (lock-key) mechanisms and to infer on
biological functionality of such interactions.

\vspace{.5cm}
\newlength{\tmp}
\setlength{\tmp}{\parindent}
\setlength{\parindent}{0pt}
{\em Acknowledgments}
\setlength{\parindent}{\tmp}

It is a pleasure to thank our collaborators I. Borukhov, J.F.
Joanny, K. Kunze, L. Leibler, R. Lipowsky, H. Orland, M. Schick
and C. Seidel with whom we have been working  on polyelectrolyte
problems. We benefitted from discussions with G. Ariel, Y.
Burak and M. Ullner. One of us (DA) would like to acknowledge partial support
from the Israel Science Foundation funded by the Israel Academy of
Sciences and Humanities --- Centers of Excellence Program and the
Israel--US Binational Science Foundation (BSF) under grant no.
98-00429.

\pagebreak
\section*{Legend of Symbols}
\begin{itemize}

\item $a$: Kuhn length (equivalent to  monomer size for flexible polymers)
\item $b$: monomer size in the case of semi-flexible polymers
\item $c_m$: monomer concentration in dilute solution.
\item $c_m(x)$: monomer profile at distance $x$ from the surface
\item $c_m^b$: bulk monomer concentration in semi-dilute
solutions.
\item $c_m^*$: overlap concentration of bulk polymer solution
\item $c_M$: characteristic value of PE adsorption concentration
(semi-dilute case)
\item $c_\s$: salt concentration in the solution
\item $c^\pm(x)$: profiles of $\pm$ ions
 \item $d$: charged rod diameter (or cross-section)
 \item $D$: adsorption layer thickness (semi-dilute case)
 \item $\delta$: thickness of adsorption layer (single chain case)
\item $e$: electronic unit charge
\item $\varepsilon$: dielectric constant of the medium.
$\varepsilon=80$ for water.
\item $f$: fraction charge of the chain $0<f<1$
\item $f^*$: crossover value for $f$
\item $f_{\rm att}$: attractive (electrostatic) adsorption force (per monomer unit length)
(single chain)
\item $f_{\rm rep}$: repulsive (entropic) adsorption force (per monomer unit length)
(single chain)
\item ${\cal F}$:  excess free energy of adsorption per unit
area (semi-dilute)
\item $F_{\rm pol}$: polymer contribution to the free energy
density (semi-dilute)
\item $F_{\rm ions}$: entropic contribution of small ions
(semi-dilute)
\item $F_{\rm el}$: electrostatic contribution (semi-dilute)
\item $F_{\rm os}$: osmotic counter-ion contribution to brush free energy
\item $F_{\rm v2}$:  second-virial contribution to brush free energy
\item $F_{\rm el}$:  direct electrostatic contribution to brush free energy
\item $F_{\rm st}$: chain stretching contribution to brush free energy
\item $\phi(x)=\sqrt{c_m(x)}$: polymer order parameter
\item $\phi_b$: bulk value of polymer order parameter
 \item $g$: number of monomer per blob
\item $\Gamma$: polymer surface excess per unit area
\item $h(x)$: dimensionless PE adsorption profile
 \item $H$: counter-ion cloud vertical extend from the surface (PE brush case)
\item $k_B T$: thermal energy
\item $\kappa^{-1}$: Debye-H\"uckel screening length
\item  $\kappa_{\rm salt}$: salt contribution to $\kappa$
 \item $L$: contour length of a chain
 \item $L$: polymer vertical extent from the surface (PE brush case)
\item $L_{\rm el}$: chain length inside one electrostatic blob
\item $L_{\rm sw}$: chain length inside one swollen blob (Sec. 3.1.2)
\item $\ell_B=e^2/(\varepsilon k_B T)$:  Bjerrum length
\item $\ell_0$: bare (mechanical) persistence length
\item $\ell_{\rm OSF}$: electrostatic contribution to persistence length (Odijk, Skolnick, and
Fixman length)
\item $\ell_{\rm OSF}^{sd}$: generalization of the OSF to semi-dilute regime.
 \item $\ell_{\rm eff}$: effective persistence length
 \item $\ell_{\rm KK}$: persistence length of a string of blobs
\item $\mu^\pm$: chemical potential of $\pm$ ions
\item $\mu_p$: chemical potential of polymer
 \item $N$: polymerization index
 \item $\nu$: Flory exponent for the polymer size
\item  $\Pi$: osmotic pressure
\item $\psi(x)$: electrostatic potential at point $x$ from the surface
\item $\psi_s$: surface potential at $x=0$
 \item $q^*$: the $q$ wavenumber at the peak of $S(q)$
 \item $Q$: total charge of particle or polyelectrolyte chain
\item  $R$: end-to-end polymer chain radius
\item $R_{\rm el}$: radius of one electrostatic blob
 \item $R_p$: spherical particle radius (Sec. 4.1)
 \item ${\bf R}_N$: end-to-end chain vector
 \item ${\bf r}_i$: position vector of the $i$th monomer
 \item $\rho$: grafting density of a PE brush
\item  $\rho^*$: crossover value for $\rho$
\item $s$: $0\le s\le L$ continuous contour chain parameter
 \item $S_0(q)$: form factor of a single chain
 \item $S(q)$: structure factor (or scattering function) of a PE solution
\item $\sigma$: surface charge density (in units of $e$) at $x=0$
 \item $\sigma^*$: minimal threshold value for $\sigma$ for single chain adsorption
\item $\Delta \sigma=f\Gamma- \sigma$: overcharging parameter
 \item $\Sigma=\sigma\ell_0^{3/2}\ell_B^{1/2}$: dimensionless adsorption parameter
\item $\tau=f/b$: linear charge density on the chain
\item $v_2$: 2nd virial coefficient
of monomers in solution. $v_2>0$ for good solvents.
 \item $v(r)=e^2/(k_B T \varepsilon r)$: dimensionless Coulomb interaction between two ions
 \item $v_{\rm DH}(r)=v(r)\exp(-\kappa r)$: Debye-H\"uckel
 interaction
 \item $v_{\rm DH}(q)$: DH interaction in Fourier space
 \item $v_{\rm RPA}(q)$: RPA interaction in Fourier space
\item $y(x)=e\psi(x)/k_BT$: dimensionless  potential profile
\item $y_s=y(0)$: rescaled surface potential
 \item $\xi$: correlation length (mesh size) of semi-dilute polymer solution
 \item $z=\pm 1$, $\pm 2, \dots$ valency of the ions
 \item $Z$: charge of the spherical particle in units of $e$ (Sec. 4.1)

\end{itemize}
\newpage


\pagebreak
\section*{Figure Captions}
\pagestyle{empty}

\bigskip
\noindent {\bf Figure~1}: Snapshots of Monte-Carlo simulations of
a neutral and semi-flexible chain consisting of $N=100$ monomers
of diameter $b$, which defines the unit of length. The theoretical
end-to-end radius $R$ is indicated by a straight bar. The
persistence lengths used in the simulations are: a) $\ell_0 = 0$,
corresponding to a freely-jointed (flexible) chain,  leading to
an end-to-end radius $ R/b = 10$, b) $\ell_0/b = 2 $, leading to
$R/b = 19.8 $, c) $\ell_0/b = 10 $, leading to $R/b = 42.4 $, d)
$\ell_0/b = 100 $, leading to $R/b = 85.8 $.

\bigskip
\noindent {\bf Figure~2}: Snapshots of  Monte-Carlo simulations of
a polyelectrolyte chain of $N=100$ monomers of size $b$, taken as
the unit length. In all simulations the bare persistence length
is fixed at $\ell_0/b =1$, and the screening length and the
charge interactions are tuned such that the electrostatic
persistence length is constant and $\ell_{\rm OSF}/b = 100 $. The
parameters used are: a) $\kappa^{-1}/b =\protect \sqrt{50}  $ and
$\tau^2 \ell_B \ell_0 =8$, b)  $\kappa^{-1}/b =\protect \sqrt{200} $
and $\tau^2 \ell_B \ell_0 =2$, c)  $\kappa^{-1}/b =\protect \sqrt{800}
$ and $\tau^2 \ell_B \ell_0 =1/2$, and d) $\kappa^{-1}/b =\protect
\sqrt{3200} $ and $\tau^2 \ell_B \ell_0 =1/8$. Noticeably, the weakly
charged chains crumple at small length scales and show a tendency
to form electrostatic blobs.

\bigskip
\noindent {\bf Figure~3}: Snapshots of Monte-Carlo simulations of
a PE chain consisting of $N=100$ monomers of size $b$. In all
simulations, the bare persistence length is fixed at $\ell_0/b
=1$, and the charge-interaction parameter is chosen to be $\tau^2
\ell_B \ell_0 =2$. The snapshots correspond to varying screening
length of: a) $\kappa^{-1}/b = \protect \sqrt{2}$, leading to an
electrostatic contribution to the persistence length of
$\ell_{\rm OSF}/b = 1$, b) $\kappa^{-1}/b =\protect \sqrt{18}$,
leading to  $\ell_{\rm OSF}/b= 9$, and c) $\kappa^{-1}/b
=\protect \sqrt{200}$, leading to $\ell_{\rm OSF}/b =100$.
According to the simple scaling principle, Eq. (\ref{elleff}), the
effective  persistence length in the snapshots a)-c) should be
similar to the bare persistence length in Figure~1b)-d).

\bigskip
\noindent {\bf Figure~4}: Schematic view of the four scaling
ranges in the Gaussian-persistent regime. On spatial scales
smaller than $R_{\rm el}$ the chain behavior is Gaussian; on
length scales larger than $R_{\rm el}$ but smaller than $\ell_{\rm
KK}$ the Gaussian blobs are aligned linearly. On length
scales up to $L_{\rm sw}$ the chain is isotropically swollen with an exponent
$\nu=1/2$, and on even larger length scales self-avoidance effects
become important and $\nu$ changes  to $\nu = 3/5$.

\bigskip
\noindent {\bf Figure~5}: Schematic phase  diagram of a single
semi-flexible PE in bulk solution with bare persistence length
$\ell_0$ and line charge density $\tau$, exhibiting various
scaling regimes. High salt concentration and small $\tau$
correspond to the Gaussian regime, where the electrostatic
interactions are irrelevant. In the persistent regime, the
polymer persistence length is increased, and in the
Gaussian-persistent regime the polymer forms a persistent chain
of Gaussian blobs as indicated in Figure~4. The broken lines
indicates the Manning condensation, at which counter-ions
condense on the polymer and reduce the effective polymer line
charge density. We use a log-log plot, and the various power-law
exponents  for the crossover boundaries are denoted by numbers.

\bigskip
\noindent {\bf Figure~6}: Schematic view of the chain structure in
the semi-dilute concentration range. The mesh size $\xi$ is about
equal to the effective polymer persistence length $\ell_{\rm eff}$
and to the screening length $\kappa^{-1}$ (if no salt is added to
the system).

\bigskip
\noindent {\bf Figure~7}: a) RPA prediction for the rescaled
structure factor $S(q)/c_m$ of a semi-dilute PE solution with
persistence length $\ell_{\rm eff}=1$\,nm, monomer length
$b=0.38$\,nm and charge fraction $f=0.5$ in the salt-free case.
The monomer densities are (from bottom to top) $c_m =$1\,M,
0.3\,M, 10\,mM, 3\,mM, 1\,mM, and 0.3\,mM. b) For the same series
of $c_m$ values as in a) the structure factor is multiplied by the
wavenumber $q$. The semi-flexibility becomes more apparent
because for large $q$ the curves tend towards a constant.

\bigskip
\noindent {\bf Figure~8}: a) Schematic picture of the adsorbed
polymer layer when  the effective persistence length is larger
than the layer thickness, $ \ell_{\rm eff} > \delta$. The distance
between two contacts of the polymer with the substrate, the
so-called deflection length, scales as $\lambda \sim \delta^{2/3}
\ell_{\rm eff}^{1/3}$.
 b) Adsorbed layer for the case when the persistence length is smaller
than the layer thickness, $\ell_{\rm eff} <\delta$. In this case
the polymer forms a random coil with many loops and a description
in terms of a flexible polymer model becomes appropriate.

\bigskip
\noindent {\bf Figure~9}: Adsorption scaling diagram shown on a
log-log plot for a) strongly charged surfaces, $\Sigma = \sigma
\ell_0^{3/2} \ell_B^{1/2} >1$ and for b) weakly charged surfaces
$\Sigma <1$. We find a desorbed regime, an adsorbed phase where
the polymer is flat and dense, and an adsorbed phase where the
polymer shows loops. It is seen that a fully charged PE is
expected to adsorb in a flat layer, whereas charge-diluted PE's
can form coiled layers with loops and dangling ends. The broken
lines denote the scaling boundaries of PE chains in the bulk as
shown in Figure~5. The numbers on the lines indicate the power law
exponents of the crossover boundaries between the regimes.

\bigskip
\noindent {\bf Figure~10}: Numerically determined adsorption
diagram for a charged semi-flexible polymer of length
$L = 50$\,nm, linear charge density $\tau = 6$\,nm$^{-1}$,
persistence length $\ell_0 = 30$\,nm, interacting with an
oppositely charged sphere of radius $R_p=5$\,nm. Shown is the main
transition from the unwrapped configuration (at the bottom) to
the wrapped configuration (at the top) as a function of sphere
charge $Z$ and inverse screening length $\kappa$. Wrapping is
favored at intermediate salt concentrations. The parameters are
chosen for the problem of DNA-histone complexation. Adapted from
ref.~\cite{Kunze}.

\bigskip
\noindent {\bf Figure~11}:
   Adsorption profiles obtained by numerical solutions of
Eqs.~(\ref{PBs}), (\ref{SCFs}) for several sets of physical
parameters in the low--salt limit. The polymer concentration
scaled by its bulk value $c_m^b=\phi_b^2$ is plotted as a function
of the distance from the surface.
  The different curves correspond to:
 $f=1$, $a=5$\AA\ and $y_s=e\psi_s/k_B T=-0.5$ (solid curve);
 $f=0.1$, $a=5$\AA\ and $y_s=-0.5$ (dots);
 $f=1$, $a=5$\AA\ and $y_s=-1.0$ (short dashes);
 $f=1$, $a=10$\AA\ and $y_s=-0.5$ (long dashes);
 and $f=0.1$,$a=5$\AA\ and $y_s=1.0$ (dot--dash line).
For all cases  $\phi_b^2=10^{-6}$\AA$^{-3}$, $v_2=50$\AA$^3$,
 $\varepsilon=80$,
 $T=300$K and $c_\s=0.1$\,mM. Adapted from ref.~\cite{mm98}.

\bigskip
\noindent {\bf Figure~12}:
  Scaling behavior of PE adsorption in the
low--salt regime Eqs.~(\ref{scalingD}), (\ref{scalingcm}). (a) The
rescaled electrostatic potential $\psi(x)/|\psi_s|$ as a function
of the rescaled distance $x/D$. (b) The rescaled polymer
concentration
 $c_m(x)/c_M$ as a function of the same rescaled distance.
The profiles are taken from Figure~11 (with the same notation).
The numerical prefactors of a linear $h(x/D)$ profile were used in
the calculation of $D$ and $c_M$. Adapted from ref.~\cite{mm98}.

\bigskip
\noindent {\bf Figure~13}:
   Typical adsorbed amount $\Gamma$ as a function of
(a) the charge fraction $f$ and (b) the ${\rm pH -pK}_0$ of the
solution for different salt concentrations
Eq.~(\ref{scalingGamma2}). The insets correspond to the low--salt
regime, Eq.~(\ref{scalingGamma1}). The parameters used for
$\varepsilon$, $T$ and $v_2$ are the same as in Figure~11, while
$y_s=e\psi_s/k_B T=-0.5$ and $a=5$\AA. The bulk concentration
$c_m^b=\phi_b^2$ is assumed to be much smaller than $c_M$.
 Adapted from ref.~\cite{mm98}.

\bigskip
\noindent {\bf Figure~14}:
   The adsorbed amount $\Gamma$ as a function of
the salt concentration $c_\s$, Eq.~(\ref{scalingGamma2}), for
$f=0.01$ and $0.25$. The solid curves on the right hand side
correspond to the scaling relations in the high--salt regime,
Eq.~(\ref{scalingGamma2}). The horizontal lines on the left hand
side mark the low salt values, Eq.~(\ref{scalingGamma1}). The
dashed lines serve as guides to the eye. The parameters used are:
 $\varepsilon=80$, $T=300$\,K, $v_2=50$\AA$^3$, $a=5$\AA, $y_s=-2.0$.
 Adapted from ref.~\cite{mm98}.

\bigskip
\noindent {\bf Figure~15}:
  Schematic diagram of the different adsorption regimes
as function of the charge fraction $f$ and the salt concentration
$c_\s$. Three regimes can be distinguished: (i) the low--salt
regime $D\ll\kappa^{-1}$; (ii) the high--salt regime (HS I)
$D\gg\kappa^{-1}$ for weak PE's
 $f\ll f^*=({c_\s v_2})^{1/2}$; and (iii) the high--salt regime (HS II)
$D\gg\kappa^{-1}$ for strong PE's $f\gg f^*$.

\bigskip
\noindent {\bf Figure~16}:
 Schematic PE brush structure. In a) we show the
weak-charge limit where the counter-ion cloud has a thickness $H$
larger than the thickness of the brush layer, $L$. In  b) we show
the opposite case of the strong-charge limit, where all
counter-ions are contained inside the brush and a single length
scale $L \approx H$  exists.

\bigskip
\noindent {\bf Figure~17}: Scaling diagram for PE brushes on a
log-log plot as a function of the grafting density $\rho$ and the
fraction of charged monomers $f$. Featured are the Pincus-brush
regime, where the counter-ion layer thickness is much larger than
the brush thickness, the osmotic-brush regime, where all
counter-ions are inside the brush and the brush height is
determined by an equilibrium between the counter-ion osmotic
pressure and the PE stretching energy, and the neutral-brush
regime, where charge effects are not important and the brush
height results from a balance of PE stretching energy and
second-virial repulsion. The power law exponents of the various
lines are denoted by numbers.

\vfil\eject

\begin{figure}[tbh]
  \epsfxsize=0.7\linewidth
  \centerline{\hbox{ \epsffile{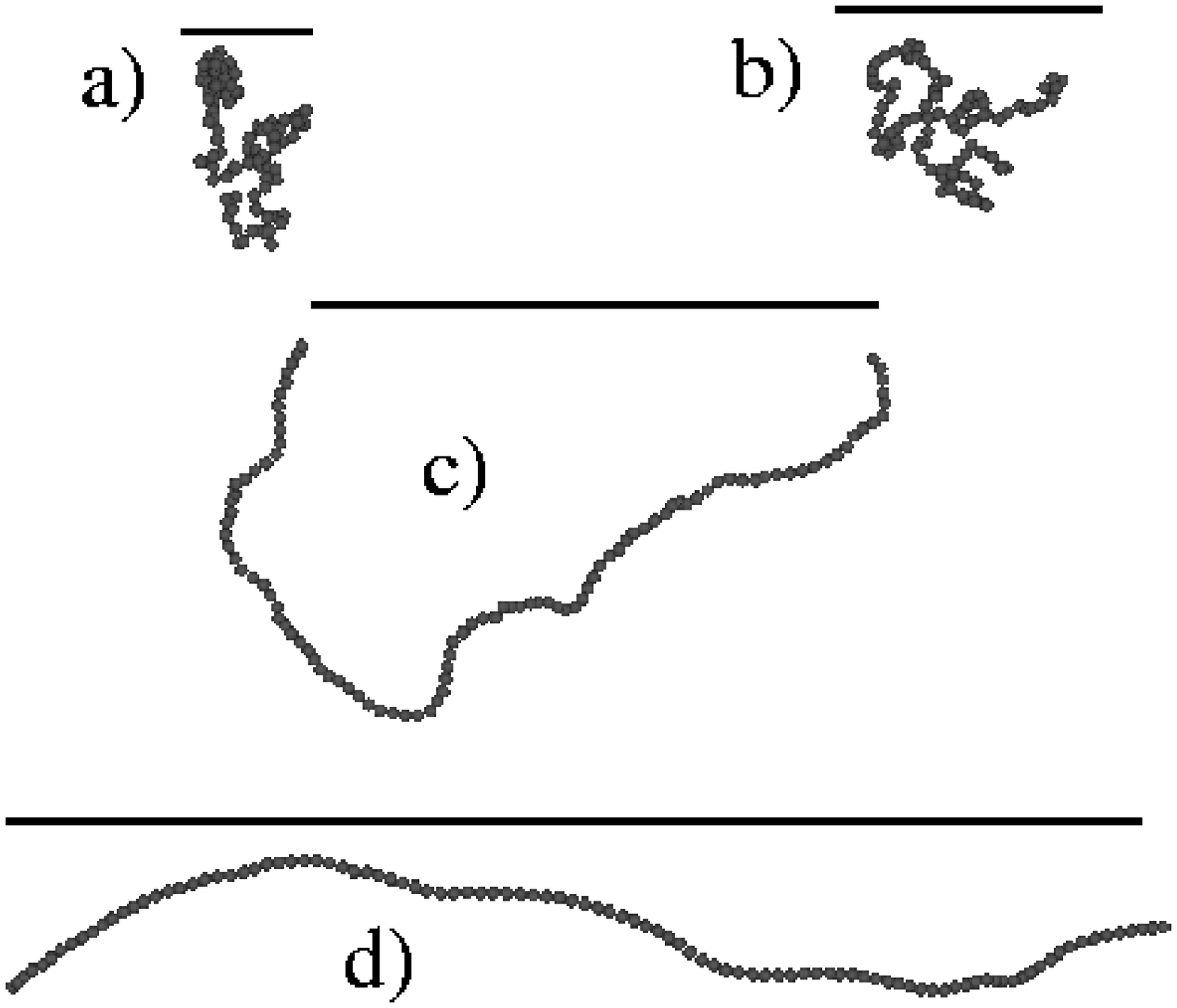}}}
\end{figure}\vfill
\vspace{2cm} {\large Netz and Andelman: Fig.~1}  \pagebreak

\begin{figure}[tbh]
  \epsfxsize=0.7\linewidth
  \centerline{\hbox{ \epsffile{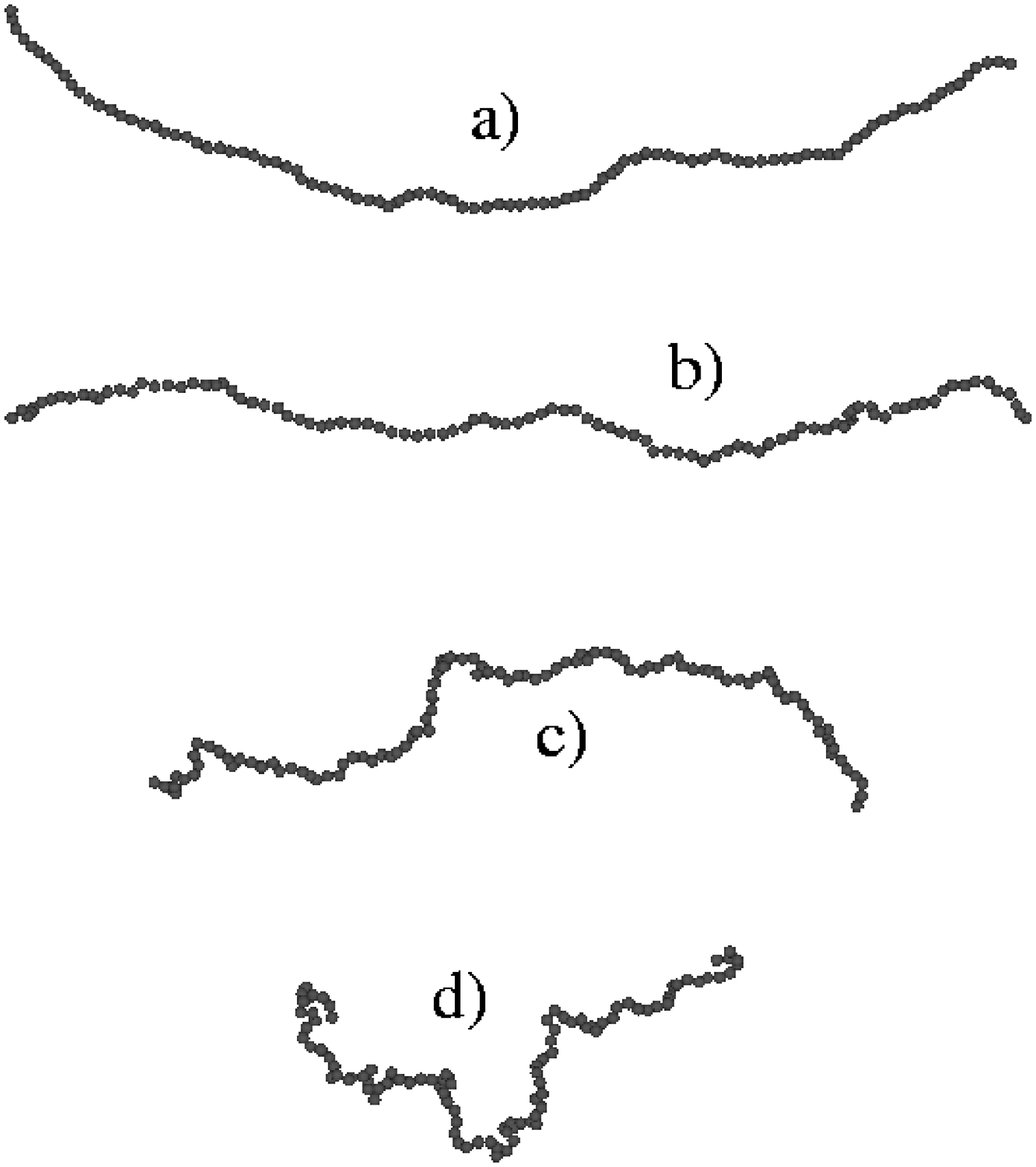}}}
\end{figure}\vfill
\vspace{3cm} {\large Netz and Andelman: Fig.~2}  \pagebreak

\begin{figure}[tbh]
  \epsfxsize=0.7\linewidth
  \centerline{\hbox{ \epsffile{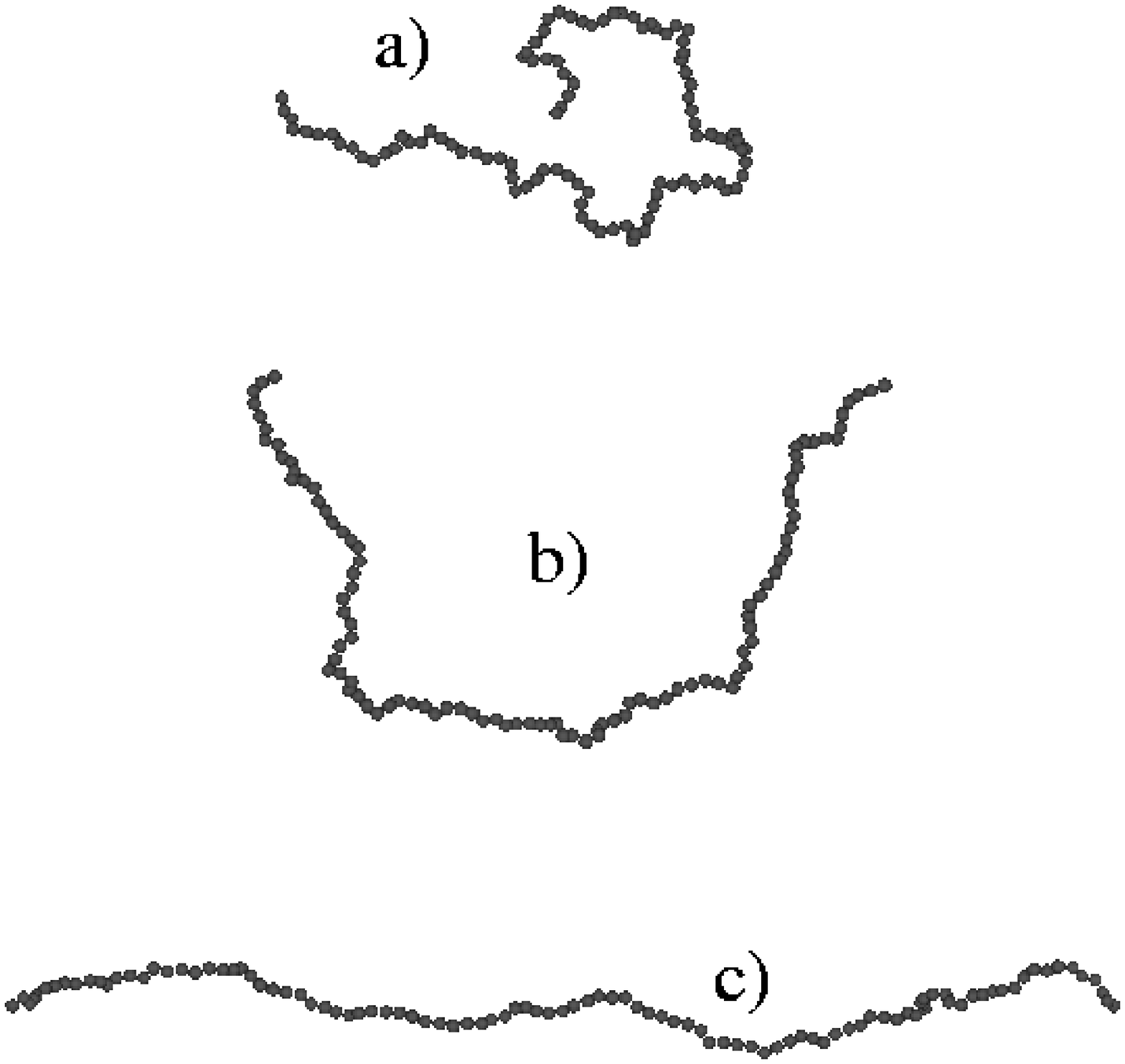}}}
\end{figure}\vfill
\vspace{3cm} {\large Netz and Andelman: Fig.~3}  \pagebreak

\begin{figure}[tbh]
  \epsfxsize=0.6\linewidth
  \centerline{\hbox{ \epsffile{fig4.ps}}}
\end{figure}\vfill
\vspace{2cm} {\large Netz and Andelman: Fig.~4}  \pagebreak

\begin{figure}[tbh]
  \epsfxsize=0.8\linewidth
  \centerline{\hbox{ \epsffile{fig5.ps}}}
\end{figure}\vfill
\vspace{3cm} {\large Netz and Andelman: Fig.~5}  \pagebreak

\begin{figure}[tbh]
  \epsfxsize=0.7\linewidth
  \centerline{\hbox{ \epsffile{fig6.ps}}}
\end{figure}\vfill
\vspace{3cm} {\large Netz and Andelman: Fig.~6}  \pagebreak

\begin{figure}[tbh]
  \epsfxsize=0.7\linewidth
  \centerline{\hbox{ \epsffile{fig7.ps}}}
\end{figure}\vfill
\vspace{4cm} {\large Netz and Andelman: Fig.~7}  \pagebreak

\begin{figure}[tbh]
  \epsfxsize=0.7\linewidth
  \centerline{\hbox{ \epsffile{fig8.ps}}}
\end{figure}\vfill
\vspace{7cm} {\large Netz and Andelman: Fig.~8}  \pagebreak

\begin{figure}[tbh]
  \epsfxsize=0.8\linewidth
  \centerline{\hbox{ \epsffile{fig9a.ps}}}
\end{figure}\vfill
\vspace{5cm} {\large Netz and Andelman: Fig.~9a}  \pagebreak

\begin{figure}[tbh]
  \epsfxsize=0.8\linewidth
  \centerline{\hbox{ \epsffile{fig9b.ps}}}
\end{figure}\vfill
\vspace{5cm} {\large Netz and Andelman: Fig.~9b}  \pagebreak

\begin{figure}[tbh]
  \epsfxsize=0.7\linewidth
  \centerline{\hbox{ \epsffile{fig10.ps}}}
\end{figure}\vfill
\vspace{7cm} {\large Netz and Andelman: Fig.~10}  \pagebreak

\begin{figure}[tbh]
  \epsfxsize=0.7\linewidth
  \centerline{\hbox{ \epsffile{fig11.ps}}}
\end{figure}\vfill
\vspace{5cm} {\large Netz and Andelman: Fig.~11}  \pagebreak

\begin{figure}[tbh]
  \epsfxsize=0.7\linewidth
  \centerline{\hbox{ \epsffile{fig12a.ps}}}
\end{figure}\vfill
\vspace{7cm} {\large Netz and Andelman: Fig.~12a}  \pagebreak

\begin{figure}[tbh]
  \epsfxsize=0.7\linewidth
  \centerline{\hbox{ \epsffile{fig12b.ps}}}
\end{figure}\vfill
\vspace{7cm} {\large Netz and Andelman: Fig.~12b}  \pagebreak

\begin{figure}[tbh]
  \epsfxsize=0.7\linewidth
  \centerline{\hbox{ \epsffile{fig13a.ps}}}
\end{figure}\vfill
\vspace{7cm} {\large Netz and Andelman: Fig.~13a}  \pagebreak

\begin{figure}[tbh]
  \epsfxsize=0.7\linewidth
  \centerline{\hbox{ \epsffile{fig13b.ps}}}
\end{figure}\vfill
\vspace{7cm} {\large Netz and Andelman: Fig.~13b}  \pagebreak

\begin{figure}[tbh]
  \epsfxsize=0.7\linewidth
  \centerline{\hbox{ \epsffile{fig14.ps}}}
\end{figure}\vfill
\vspace{7cm} {\large Netz and Andelman: Fig.~14}  \pagebreak

\begin{figure}[tbh]
  \epsfxsize=0.7\linewidth
  \centerline{\hbox{ \epsffile{fig15.ps}}}
\end{figure}\vfill
\vspace{3cm} {\large Netz and Andelman: Fig.~15}  \pagebreak

\begin{figure}[tbh]
  \epsfxsize=0.5\linewidth
  \centerline{\hbox{ \epsffile{fig16.ps}}}
\end{figure}\vfill
\vspace{3cm} {\large Netz and Andelman: Fig.~16}  \pagebreak

\begin{figure}[tbh]
  \epsfxsize=0.5\linewidth
  \centerline{\hbox{ \epsffile{fig17.ps}}}
\end{figure}\vfill
\vspace{3cm} {\large Netz and Andelman: Fig.~17}  \pagebreak

\end{document}